\documentstyle[12pt,aps,prd,preprint]{revtex}
\tightenlines

\begin{document}

\title{TRUE TRANSFORMATIONS OF SPACETIME LENGTHS AND\ \ APPARENT\ TRANSFORMATIONS\
OF\ SPATIAL\ AND\ TEMPORAL\ DISTANCES. II. THE COMPARISON WITH EXPERIMENTS }
\author{Tomislav Ivezi\'{c} \\
\textit{Ru{%
\mbox
 {\it{d}\hspace{-.15em}\rule[1.25ex]{.2em}{.04ex}\hspace{-.05em}}}er Bo\v
{s}kovi\'{c} Institute, P.O.B. 180, 10002 Zagreb, Croatia}\\
\textit{\ ivezic@rudjer.irb.hr}}
\maketitle

Some of the well-known experiments: the ''muon'' experiment, the
Michelson-Morley type experiments, the Kennedy-Thorndike type experiments
and the Ives-Stilwell type experiments are analyzed using the
nonrelativistic theory, the ''apparent transformations (AT) relativity'' and
the ''true transformations (TT) relativity.'' It is shown that all the
experiments (when they are complete from the ''TT relativity'' viewpoint)
are in agreement with the ''TT relativity'' in which the special relativity
is understood as the theory of a four-dimensional spacetime with the
pseudo-Euclidean geometry. It is also explicitly shown that, in contrast to
the usual opinion, the commonly used ''AT relativity'' does not always agree
with experiments. The concept of sameness of a physical quantity is
essential for the distinction between the two forms of relativity both in
the theory and in experiments. The difference in this concept causes the
agreement of the ''TT relativity'' with the experiments and the disagreement
of the ''AT relativity.''\medskip

\noindent PACS number(s): 03.30.+p\pagebreak

\noindent \textit{Henceforth space by itself, and time by itself, are doomed 
\newline
to fade away into mere shadows and only a kind of union of \newline
the two will preserve an independent reality.} H. Minkowski\medskip

\noindent \textit{A quantity is therefore physically meaningful (in the
sense that it is of the same nature to all observers) if it has tensorial
properties under Lorentz transformations. }F. Rohrlich

\section{INTRODUCTION}

In \cite{ivezic}, \cite{ive2} and \cite{ivtrue} (this paper will be referred
as $\left[ I\right] $), (see also \cite{ivsc98}), two forms of relativity
are discussed, the ''true transformations (TT) relativity'' and the
''apparent transformations (AT) relativity.'' The notions of the TT and the
AT are first introduced by Rohrlich \cite{rohrl1}, and, in the same meaning,
but not under that name, discussed in \cite{gamba} too. The general
theoretical discussion of the difference between the ''TT relativity'' and
the ''AT relativity'' is given in detail in $\left[ I\right] $. There (in $%
\left[ I\right] $) we have also presented the theoretical discussion of the
TT of the spacetime length for a moving rod and a moving clock, and of the
AT for the same examples, i.e., the AT of the spatial distance, the Lorentz
''contraction,'' and the AT of the temporal distance, the time
''dilatation.'' In this paper we use theoretical results from $\left[
I\right] $ and compare them with some experimental results.

\section{GENERAL DISCUSSION\ OF\ THE COMPARISON}

It is usually interpreted that the experiments on ''length contraction'' and
''time dilatation'' test the special relativity, but the discussion from $%
\left[ I\right] $ shows that such an interpretation of the experiments
refers to - the ''AT relativity,'' and not to - the ''TT relativity.''

It has to be noted that in the experiments in the ''TT relativity,'' in the
same way as in the theory, see $\left[ I\right] $, the measurements in
different inertial frames of reference (IFRs) (and different
coordinatizations) have to refer to the same four-dimensional (4D) tensor
quantity. In the chosen IFR and the chosen coordinatization the measurement
of some 4D quantity has to contain the measurements of all parts of such a
quantity. However in almost all experiments that refer to the special
relativity only the quantities belonging to the ''AT relativity'' were
measured. From the ''TT relativity'' viewpoint such measurements are
incomplete, since only some parts of a 4D quantity, not all, are measured.
This fact presents a serious difficulty in the reliable comparison of the
existing experiments with the ''TT relativity,'' and, actually, we shall be
able to compare in a quantitative manner only some of the existing
experiments with the ''TT relativity.''

To examine the differences between the nonrelativistic theory, the commonly
used ''AT relativity,'' and the ''TT relativity'' we shall make the
comparison of these theories with some experiments in the following sections.

\section{THE ''MUON'' EXPERIMENT}

First we shall examine an experiment in which different results will be
predicted for different synchronizations in the conventional approach to
relativity, i.e., in the ''AT relativity,'' but of course the same results
for all synchronizations will be obtained in the ''TT relativity.'' This is
the ''muon'' experiment, which is already theoretically discussed from the
''TT relativity'' viewpoint in Sec. 3.2 in $\left[ I\right] $ and from the
''AT relativity'' viewpoint in Sec. 4.2 in $\left[ I\right] .$ The ''muon''
experiment is quoted in almost every text-book on general physics, see,
e.g., \cite{Feyn} and \cite{Kittel}. Moreover, an experiment \cite{Frisch}
was the basis for a film often shown in introductory modern physics courses:
''Time dilation: An experiment with $\mu $ mesons.'' Recently, in \cite
{Easwar}, a version of such an experiment is presented, and it required
travelling to mountains of moderate heights of around 600 m.

In these experiments, \cite{Frisch} and \cite{Easwar}, the fluxes of muons
on a mountain, $N_{m}$, and at sea level, $N_{s}$, are measured, and the
number of muons which decayed in flight is determined from their difference.
Also the distribution of the decay times is measured for the case when the
muons are at rest, giving a lifetime $\tau $ of approximately $2.2\mu s.$
The rate of decay of muons at rest, i.e., in the muon frame, is compared
with their rate of decay in flight, i.e., in the Earth frame. In \cite
{Frisch} high-velocity muons are used, which causes that the fractional
energy loss of the muons in the atmosphere is negligible, making it a
constant velocity problem, while in \cite{Easwar} one deals with a variable
velocity problem. The discussion of the ''muon'' experiment in $\left[
I\right] $ referred to the decay of only one particle. When the real
experiments are considered, as are \cite{Frisch} and \cite{Easwar}, then we
use data on the decay of many such radioactive particles and the
characteristic quantities are avareged over many single decay events.

\subsection{The nonrelativistic approach}

In the nonrelativistic theory the space and time are separated. The
coordinate transformations connecting the Earth frame and the muon frame are
the Galilean transformations giving that $t_{E}$, the travel time from the
mountain to sea level when measured in the Earth frame, is the same as $%
t_{\mu }$, which is the elapsed time for the same travelling but measured in
the moving frame of the muon, $t_{E}=t_{\mu }$. Also, in the nonrelativistic
theory, the lifetimes of muons in the mentioned two frames are equal, $\tau
_{E}=\tau _{\mu }=\tau .$ Muon counts on the mountain $N_{m},$ and at sea
level $N_{s},$ as experimentally determined numbers, must not depend on the
frame in which they are measured and on the chosen coordinatization. This
result, i.e., that $N_{s\mu }$=$N_{sE}=N_{s}$ and $N_{m\mu }=N_{mE}=N_{m},$
has to be obtained not only in the nonrelativistic theory but also in the
''AT relativity'' and in the ''TT relativity.'' The differential equation
for the radioctive-decay processes in the nonrelativistic theory can be
written as 
\begin{equation}
dN/dt=-\lambda N,\quad N_{s}=N_{m}\exp (-t/\tau ).  \label{radclas}
\end{equation}
The travel time $t_{E}$ is not directly measured by clocks, than, in the
Earth frame, it is determined as the ratio of the height of the mountain $%
H_{E}$ and the velocity of the muons $v$, $t_{E}=H_{E}/v.$ The equation (\ref
{radclas}) holds in the Earth frame and in the muon frame too, since the two
frames are connected by the Galilean transformations, and, as mentioned
above, the corresponding times are equal, $t_{E}=t_{\mu }$ and $\tau
_{E}=\tau _{\mu }.$ Hence we conclude that in the nonrelativistic theory the
exponential factors are the same in both frames and consequently the
corresponding fluxes in the two frames are equal, $N_{s\mu }$=$N_{sE}$ and $%
N_{m\mu }=N_{mE}$, as it must be. However the experiments show that the
actual flux at sea level is much higher than that expected from such a
nonrelativistic calculation, and thus the nonrelativistic theory does not
agree with the experimental results.

\subsection{The ''AT relativity'' approach}

In the ''AT relativity'' different physical phenomena in different IFRs must
be invoked to explain the measured values of the fluxes; the time
''dilatation'' is used in the Earth frame, but in the muon frame one
explains the data by means of the Lorentz ''contraction.'' In order to
exploit the results of Secs. 3.2 and 4.2 in $\left[ I\right] $ we analyse
the ''muon'' experiment not only in the ''e'' coordinatization but also in
the ''r'' coordinatization. As shown in Sec.4 in $\left[ I\right] $ the ''AT
relativity'' considers that the spatial and temporal parts of the spacetime
length are well-defined physical quantities in 4D spacetime. (But, of
course, Sec.4 in $\left[ I\right] $ also reveals that such an assumption
holds only in the Einstein coordinatization, i.e., in the ''e'' base, see $%
\left[ I\right] $.)

Then, as in the nonrelativistic theory, the equation for the
radioactive-decay in the ''AT relativity'' can be written as 
\begin{equation}
dN/dx^{0}=-\lambda N,\quad N_{s}=N_{m}\exp (-\lambda x^{0}).  \label{radate}
\end{equation}
The equation (\ref{radate}) contains a specific coordinate, the $x^{0}$
coordinate, which means that the equation (\ref{radate}) will not remain
unchanged upon the Lorentz transformation, i.e., it will not have the same
form in different IFRs (and also in different coordinatizations). But in the
''AT relativity'' it is not required that the physical quantities must be
the 4D tensor quantities that correctly transform upon the Lorentz
transformations. Thus the quantities in (\ref{radate}) are not the 4D tensor
quantities, which actually causes that different phenomena in different IFRs
have to be invoked to explain the same physical effect, i.e., the same
experimental data. In the Earth frame and in the ''e'' base we can write in (%
\ref{radate}) that $x_{E}^{0}=ct_{E},$ $\lambda _{E}=1/c\tau _{E},$ which
gives that the radioactive-decay law becomes $N_{sE}=N_{mE}\exp (-t_{E}/\tau
_{E}).$ In the experiments \cite{Frisch} and \cite{Easwar} $N_{sE},$ $%
N_{mE}, $ and $t_{E}=H_{E}/v$ are measured in the Earth frame (tacitly
assuming the ''e'' coordinatization). However the lifetime of muons is
measured in their rest frame. Now, in contrast to the nonrelativistic theory
where $\tau _{E}=\tau _{\mu }$ and $t_{E}=t_{\mu },$ the ''AT relativity''
assumes that in the ''e'' base there is the time ''dilatation'' determined
by Eq.(20) in Sec.4.2 in $\left[ I\right] $, which gives the connection
between the lifetimes of muons in the Earth frame $\tau _{E}$ and the
measured lifetime in the muon frame $\tau _{\mu }$ as 
\begin{equation}
\tau _{E}=\gamma \tau _{\mu }.  \label{taue}
\end{equation}
Using that relation one finds that the radioactive-decay law, when expressed
in terms of the measured quantities, becomes 
\begin{equation}
N_{sE}=N_{mE}\exp (-t_{E}/\tau _{E})=N_{mE}\exp (-t_{E}/\gamma \tau _{\mu }).
\label{radAT1}
\end{equation}
This equation is used in \cite{Frisch} to make the ''relativistic''
calculation and compare it with the experimental data. In fact, in \cite
{Frisch}, the comparison is made between the predicted time dilatation
factor $\gamma $ of the muons and an observed $\gamma .$ The predicted $%
\gamma $ is $8.4\pm 2,$ while the observed $\gamma $ is found to be $8.8\pm
0.8$, which is a convincing agreement. The prediction of $\gamma $ is made
from the measured energies of muons on the mountain and at sea level; these
energies are determined from the measured amount of material which muons
penetrated when stopped, and then the energies are converted to the speeds
of the muons using the relativistic relation between the total energy and
the speed. The observed $\gamma $ is determined from the relation (\ref
{radAT1}), where the measured rates were $N_{sE}=397\pm 9$ and $%
N_{mE}=550\pm 10,$ and the measured height of the mountain is $H_{E}=1907m.$
The lifetime of muons $\tau _{\mu }$ in the muon frame is taken as the
information from other experiments (in order to obtain more accurate result)
and it is $\tau _{\mu }=2.211\cdot 10^{-6}s.$ In \cite{Easwar} the relation
for the ''relativistic'' calculation is written as $N_{s}=N_{m}\exp (-t_{\mu
}/\tau _{\mu })=N_{m}\exp (-t_{E}/(\gamma \tau _{\mu })),$ Eq.(2) in \cite
{Easwar}, which shows that the time dilatation is taken into account by the
relation $t_{\mu }=t_{E}/\gamma $ (the same can be concluded from Eqs.(6)
and (7) in \cite{Easwar}). For a given measured flux $N_{sE}$ of muons at
sea level, $N_{sE}=95\pm 10,$ the expected flux on the mountain is
determined from a nonrelativistic calculation, Eq.(\ref{radclas}), and from
a ''relativistic'' calculation, Eq.(\ref{radAT1}), i.e., their Eq.(2). The
comparison is made between these expected fluxes and the measured counts on
the mountain. The predicted counts on the mountain ignoring time dilatation
(from (\ref{radclas})) $=330\pm 60;$ predicted counts on the mountain taking
into account time dilatation (from (\ref{radAT1})) $=190\pm 20;$ measured
counts on the mountain $=183,$ and different error bars are reported for
this measured counts. We see that the nonrelativistic calculation does not
agree with the experimentally found numbers, while the ''AT relativity''
calculation (made in the ''e'' base) shows a very convincing agreement with
measured fluxes.

Let us now see how the experiments are interpreted in the muon frame. (We
note that both \cite{Frisch} and \cite{Easwar} compared the theory (the ''AT
relativity'') and the experiments only in the Earth frame, but using $\tau
_{\mu }$ from the muon frame.) First we have to find the form of the law for
the radioactive-decay processes (\ref{radate}) in the muon frame. As
considered above the radioactive-decay law $N_{sE}=N_{mE}\exp (-t_{E}/\tau
_{E})$ in the Earth frame and in the ''e'' base is obtained from the
equation (\ref{radate}) using the relations $x_{E}^{0}=ct_{E}$ and $\lambda
_{E}=1/c\tau _{E}.$ But, as already said, the equation (\ref{radate}) does
not remain unchanged upon the Lorentz transformation and accordingly it
cannot have the same form in the Earth frame and in the muon frame. So,
actually, in the 4D spacetime, the equation for the radioactive-decay
processes in the muon frame could have, in principle, a different functional
form than the equation (\ref{radAT1}), which describes the same radioactive-
decay processes in the Earth frame. However, in the ''AT relativity,''
despite of the fact that the quantities in the Earth frame and in the muon
frame are not connected by the Lorentz transformations, the equation for the
radioactive-decay processes in the muon frame is obtained from the equation (%
\ref{radate}) in the same way as in the Earth frame, i.e., writting that $%
x_{\mu }^{0}=ct_{\mu },$ and $\lambda _{\mu }=1/c\tau _{\mu },$ (as seen in
Eq.(2) in \cite{Easwar}, the relation for the ''relativistic'' calculation),
whence 
\begin{equation}
N_{s\mu }=N_{m\mu }\exp (-t_{\mu }/\tau _{\mu }).  \label{radmu}
\end{equation}
The justification for such a procedure can be done in the following way. In
the ''AT relativity'' the principle of relativity acts as some sort of
''Deus ex machina,'' which resolves problems; the relation (\ref{radate}) is 
\emph{proclaimed} to be the physical law and the principle of relativity
requires that a physical law must have the same form in different IFRs.
(This is the usual way in which the principle of relativity is understood in
the ''AT relativity.'') Therefore, one can write in the equation (\ref
{radate}) that $x_{E}^{0}=ct_{E}$ and $\lambda _{E}=1/c\tau _{E}$ in the
Earth frame and $x_{\mu }^{0}=ct_{\mu },$ and $\lambda _{\mu }=1/c\tau _{\mu
}$ in the muon frame. With such substitutions the form of the law is the
same in both frames, as it is required by the principle of relativity. Then,
as we have already seen, when the consideration is done in the Earth frame,
the relation (\ref{taue}) for the time dilatation is used to connect
quantities in two frames,.instead of to connect them by the Lorentz
transformations. When the consideration is performed in the muon frame
another relation is invoked to connect quantities in two frames. Namely it
is considered in the ''AT relativity'' that in the muon frame the mountain
is moving and the muon ''sees'' the height of the mountain Lorentz
contracted, 
\begin{equation}
H_{\mu }=H_{E}/\gamma ,  \label{hacon}
\end{equation}
which is Eq.(18), Sec.4.1 in $\left[ I\right] ,$ for the Lorentz
contraction, giving that 
\begin{equation}
t_{\mu }=H_{\mu }/v=H_{E}/\gamma v=t_{E}/\gamma .  \label{hami}
\end{equation}
This leads to the same exponential factor in Eq.(\ref{radmu}) as that one in
the Earth frame in Eq.(\ref{radAT1}), $\exp (-t_{\mu }/\tau _{\mu })=\exp
(-t_{E}/(\gamma \tau _{\mu })).$ From that result it is concluded that in
the ''AT relativity'' and in the ''e'' base the corresponding fluxes are
equal in the two frames, $N_{s\mu }$=$N_{sE}=N_{s}$ and $N_{m\mu
}=N_{mE}=N_{m}.$ Strictly speaking, it is not the mentioned equality of
fluxes, but the equality of ratios of fluxes, $N_{sE}/N_{mE}=$ $N_{s\mu
}/N_{m\mu }$, which follows from the equality of the exponential factors in (%
\ref{radAT1}) and (\ref{radmu}). Both \cite{Frisch} and \cite{Easwar}
compared the theory (the ''AT relativity'') and the experiments only in the
Earth frame, but using $\tau _{\mu }$ from the muon frame. In \cite{Frisch}
the time $t_{\mu }$ that the muons spent in flight according to their own
clocks was inferred from the measured distribution of decay times of muons
at rest, and in \cite{Easwar} $t_{E}$ and $t_{\mu }$ are calculated by a
simple computer program using the known relation for the mean rate of energy
loss of the muons as they travel from the mountain to the sea level and
dissipate their energy in the medium (such program is necessary since in 
\cite{Easwar} one deals with a variable velocity problem).) Since the
predicted fluxes $N_{sE}$ and $N_{mE}$ are in a satisfactory agreement with
the measured ones, and since the theory (which deals with the time
dilatation and the Lorentz contraction) predicts their independence on the
chosen frame, it is generally accepted that the ''AT relativity'' correctly
explains the measured data.

The above comparison is worked out only in the ''e'' coordinatization, but
the physics demands that the independence of the fluxes on the chosen frame
must hold in all coordinatizations. Therefore we now discuss the experiments 
\cite{Frisch} and \cite{Easwar} from the point of view of the ''AT
relativity'' but in the ''r'' coordinatization, see $\left[ I\right] $.
Then, using Eq.(\ref{radate}), we can write the relation for the fluxes in
the ''r'' base and in the Earth frame as 
\[
N_{r,sE}=N_{r,mE}\exp (-\lambda _{r,E}x_{r,E}^{0})=N_{r,mE}\exp
(-x_{r,E}^{0}/x_{r,E}^{0}(\tau _{E})),
\]
where $x_{r,E}^{0}(\tau _{E})=1/\lambda _{r,E}.$ Again, as in the ''e''
base, we have to express $x_{r,E}^{0}(\tau _{E})$ in the Earth frame in
terms of the measured quantity $x_{r,\mu }^{0}(\tau _{\mu })$ using the
relation (21) from $\left[ I\right] $ for the time ''dilatation'' in the
''r'' base, 
\[
x_{r,E}^{0}(\tau _{E})=(1+2\beta _{r})^{1/2}c\tau _{\mu }.
\]
Hence, the radioactive-decay law (\ref{radate}), in the ''r'' base, and when
expressed in terms of the measured quantities, becomes 
\begin{equation}
N_{r,sE}=N_{r,mE}\exp (-x_{r,E}^{0}/(1+2\beta _{r})^{1/2}c\tau _{\mu }),
\label{radiner}
\end{equation}
and it corresponds to the relation (\ref{radAT1}) in the ''e'' base. If we
express $\beta _{r}$ in terms of $\beta =v/c$ as $\beta _{r}=\beta /(1-\beta
)$ (see $\left[ I\right] $), and use Eq.(8) from $\left[ I\right] $ to
connect the ''r'' and ''e'' bases, $%
x_{r,E}^{0}=x_{E}^{0}-x_{E}^{1}=ct_{E}-H_{E},$ then the exponential factor
in Eq.(\ref{radiner}) becomes $=\exp \left\{ -(ct_{E}-H_{E})/\left[ (1+\beta
)/(1-\beta )\right] ^{1/2}c\tau _{\mu }\right\} .$ Using $H_{E}=vt_{E}$ this
exponential factor can be written in the form that resembles to that one in (%
\ref{radAT1}), i.e., it is $=\exp (-t_{E}/\Gamma _{rE}\tau _{\mu }),$ and
Eq.(\ref{radiner}) can be written as 
\begin{equation}
N_{r,sE}=N_{r,mE}\exp (-t_{E}/\Gamma _{rE}\tau _{\mu }).  \label{raderAT}
\end{equation}
We see that $\gamma =(1-\beta )^{-1/2}$ in (\ref{radAT1}) (the ''e'' base)
is replaced by a different factor 
\begin{equation}
\Gamma _{rE}=(1+\beta )^{1/2}(1-\beta )^{-3/2}=(1+\beta )(1-\beta
)^{-1}\gamma   \label{gama}
\end{equation}
in (\ref{raderAT}) (the ''r'' base). The observed $\Gamma _{rE}$ in the
experiments \cite{Frisch} must remain the same, $=8.8\pm 0.8,$ (it is
determined from (\ref{raderAT}) with the measured values of $%
N_{r,sE},N_{r,mE},t_{E}$ and $\tau _{\mu }$), but the predicted $\Gamma
_{rE},$ using the above relation for $\Gamma _{r}$ and the known, predicted, 
$\gamma =8.4\pm 2,$ becomes $\simeq 250\gamma ,$%
\begin{equation}
\Gamma _{rE}\simeq 250\gamma .  \label{gamaer}
\end{equation}
We see that from the common point of view a quite unexpected result is
obtained in the ''r'' coordinatization; the observed $\Gamma _{rE}$ is as
before $=8.8,$ while the predicted $\Gamma _{rE}$ is $\simeq 250\cdot
8.4=2100.$ Similarly, one can show that there is a great discrepancy between
the fluxes measured in \cite{Frisch} and \cite{Easwar} and the fluxes
predicted when the ''dilatation'' of time is taken into account but in the
''r'' coordinatization. Furthermore, it can be easily proved that predicted
values in the ''r'' base and in the muon frame will again greatly differ
from the measured ones. \emph{Such results explicitly show that the ''AT
relativity'' is not a satisfactory relativistic theory; it predicts, e.g.,
different values of the flux }$N_{s}$ \emph{(for the same measured} $N_{m}$%
\emph{)} \emph{in different synchronizations and for some synchronizations
these predicted values are quite different than the measured ones. }These
results are directly contrary to the generally accepted opinion about the
validity of the ''AT relativity.''

\subsection{The ''TT relativity'' approach}

Let us now examine the experiments \cite{Frisch} and \cite{Easwar} from the
point of view of the ''TT relativity.'' In the ''TT relativity'' all
quantities entering into physical laws must be 4D tensor quantities, and
thus with correct transformation properties; \emph{the same 4D quantity} has
to be considered in different IFRs and different coordinatizations. In the
usual, ''AT relativity,'' analysis of the ''muon'' experiment, for example,
the lifetimes $\tau _{E}$ and $\tau _{\mu }$ are considered as the same
quantity. Although the transformation connecting $\tau _{E}$ and $\tau _{\mu
}$ (the dilatation of time, Eq.(\ref{taue})) is only \emph{a part} of the
Lorentz transformation written in the ''e'' base, it is believed by all
proponents of the ''AT relativity'' that $\tau _{E}$ and $\tau _{\mu }$
refer to the same temporal distance (the same quantity) but measured by the
observers in two relatively moving IFRs. However, as shown in the preceding
sections and in $\left[ I\right] ,$see Fig.4, in 4D spacetime $\tau _{E}$
and $\tau _{\mu }$ refer to different quantities, which are not connected by
the Lorentz transformation. To paraphrase Gamba \cite{gamba}: ''As far as
relativity is concerned, quantities like $\tau _{E}$ and $\tau _{\mu }$ are
different quantities, not necessarily related to one another. To ask the
relation between $\tau _{E}$ and $\tau _{\mu }$ from the point of view of
relativity, is like asking what is the relation between the measurement of
the radius of the Earth made by an observer $S$ and the measurement of the
radius of Venus made by an observer $S^{\prime }.$ We can certainly take the
ratio of the two measures; what is wrong is the tacit assumption that
relativity has something to do with the problem just because the
measurements were made by \emph{two} observers.''

Hence, in the ''TT relativity,'' instead of the equation (\ref{radate}),
which contains $x^{0}$ coordinate, we formulate the radioactive-decay law in
terms of covariantly defined quantities 
\begin{equation}
dN/dl=-\lambda N,\quad N=N_{0}\exp (-\lambda l).  \label{radTT}
\end{equation}
$l$ is the spacetime length (defined by Eq.(2) in $\left[ I\right] $; in the
abstract index notation $l=(l^{a}g_{ab}l^{b})^{1/2},$ where $l^{a}(l^{b})$
is the distance 4-vector between two events $A$ and $B$, $%
l^{a}=l_{AB}^{a}=x_{B}^{a}-x_{A}^{a}$, $x_{A,B}^{a}$ are the position
4-vectors and $g_{ab}$ is the metric tensor) for the events of creation of
muons (here on the mountain; we denote it as the event $O$) and their
arrival (here at sea level; the event $A$). $\lambda =1/l(\tau );$ $l(\tau )$
is the spacetime length for the events of creation of muons (here on the
mountain; the event $O$) and their decay after the lifetime $\tau ,$ the
event $T$. $l,$ defined in such a way, i.e., as a geometrical quantity, is
invariant upon the covariant 4D Lorentz transformations (Eq.(3) in $\left[
I\right] $)$;$ 
\[
L^{a}{}_{b}\equiv L^{a}{}_{b}(v)=g^{a}{}_{b}-\frac{2u^{a}v_{b}}{c^{2}}+\frac{%
(u^{a}+v^{a})(u_{b}+v_{b})}{c^{2}-u\cdot v}, 
\]
where $u^{a}$ is the proper velocity 4-vector of a frame $S$ with respect to
itself and $v^{a}$ is the proper velocity 4-vector of $S^{\prime }$ relative
to $S$) and, as $l$ is written in the abstract index notation, it does not
depend on the chosen coordinatization in the considered IFR. Then in the
''e'' base and in the muon frame the distance 4-vector $l_{OA}^{a}$ becomes $%
l_{\mu ,OA}^{\alpha }=(ct_{\mu },0)$ (the subscript $\mu $ will be used, as
previously in this section, to denote the quantities in the muon frame,
while Greek indices $\alpha ,\beta $ denote the components of some geometric
object, e.g., the components $l_{\mu ,OA}^{\alpha }$ in the muon frame of
the distance 4-vector $l_{OA}^{a},$ see $\left[ I\right] $ for the notation)
and the spacetime length $l$ between these events is $l_{OA}=(l_{\mu
,OA}^{\beta }l_{\mu ,\beta OA})^{1/2}=(-c^{2}t_{\mu }^{2})^{1/2}.$ The
representation of the distance 4-vector $l_{OT}^{a}$ in the ''e'' base and
in the muon frame is $l_{\mu ,OT}^{\alpha }=(c\tau _{\mu },0),$ whence the
spacetime length $l_{OT}=(l_{\mu ,OT}^{\beta }l_{\mu ,\beta
OT})^{1/2}=(-c^{2}\tau _{\mu }^{2})^{1/2}.$ Inserting the spacetime lengths $%
l_{OA}$ and $l_{OT}$ into the equation (\ref{radTT}) we find the expression
for the radioactive-decay law in the ''TT relativity'' 
\begin{equation}
N_{s}=N_{m}\exp (-l_{OA}/l_{OT}),  \label{radlaw}
\end{equation}
which in the ''e'' base and in the muon frame takes the same form as the
relation (\ref{radmu}) (the radioactive-decay law in the ''AT relativity''
in the ''e'' base and in the muon frame), 
\begin{equation}
N_{s}=N_{m}\exp (-l_{OA}/l_{OT})=N_{m}\exp (-t_{\mu }/\tau _{\mu }).
\label{radla1}
\end{equation}
Since the spacetime length $l$ is independent on the chosen IFR and on the
chosen coordinatization the relation (\ref{radlaw}) holds in the same form
in the Earth frame and in the muon frame and in both coordinatizations, the
''e'' and ''r'' coordinatizations. Hence we do not need to examine Eq.(\ref
{radlaw}) in the Earth frame, and in the ''r'' base, but we can simply
compare the relation (\ref{radla1}) with the experiments.

Thus, taking into account the discussion given at the begining of Sec.4 in $%
\left[ I\right] $, we conclude that, in order to check the validity of the
''TT relativity'' in the ''muon'' experiment, we would need, strictly
speaking, to measure, e.g., the lifetime $\tau _{\mu }$ and the time $t_{\mu
}$ in the muon frame, where they determine $l_{OT}$ and $l_{OA}$
respectively, and then to measure \emph{the same events} (that determined $%
\tau _{\mu }$ and $t_{\mu }$ in the muon frame) in an IFR that is in uniform
motion relative to the muon frame (at us it is the Earth frame). Of course
it is not possible to do so in the real ''muon'' experiment but,
nevertheless, in this case we can use the data from experiments \cite{Frisch}
and \cite{Easwar} and interpret them as that they were obtained in the way
required by the ''TT relativity.'' The reasons for such a conclusion are the
identity of microparticles of the same sort, the assumed homogeneity and
isotropy of the spacetime, and some other reasons that are actually
discussed in \cite{Frisch} (although from another point of view). Here we
shall not discuss this, in principle, a very complex question, than we take
the measured values of $\tau _{\mu },$ $t_{\mu },$ $N_{s}$ and $N_{m}$ and
compare them with the results predicted by the relation (\ref{radla1}). In 
\cite{Frisch} $\tau _{\mu }$ is taken to be $\tau _{\mu }=2.211\mu s,$ $%
N_{s}=397\pm 9,$ $N_{m}=550\pm 10,$ but $t_{\mu }$ is not measured than it
is estimated from Fig.6(a) in \cite{Frisch} to be $t_{\mu }=0.7\mu s.$
Inserting the values of $\tau _{\mu },$ $t_{\mu }$ and $N_{m}$ from \cite
{Frisch} (for this simple comparison we take only the mean values without
errors) into Eq.(\ref{radla1}) we predict that $N_{s}$ is $N_{s}=401,$ which
is in an excellent agreement with the measured $N_{s}=397.$ As it is already
said, the spacetime length $l$ takes the same value in both frames and both
coordinatizations, $l_{e,\mu }=l_{e,E}=l_{r,\mu }=l_{r,E}.$ Hence, for the
measured $N_{m}=550$ and if the distance 4-vectors $l_{OA}^{\alpha }$ and $%
l_{OT}^{\alpha }$ would be measured in the Earth frame, and in both frames
in the ''r'' base, we would find the same $N_{s}=401.$ This result
undoubtedly confirms the consistency and the validity of the ''TT
relativity.'' (Note that we cannot compare the experiments \cite{Easwar}
with the ''TT relativity'' since their $t_{\mu },$ Eq.(7) in \cite{Easwar},
is not correctly determined from the ''TT relativity'' viewpoint.)

\emph{The nonrelativistic theory predicts the same value of the exponential
factor in both frames, }$\exp (-t_{E}/\tau _{E})=\exp (-t_{\mu }/\tau _{\mu
}),$\emph{\ since it deals with the absolute time, i.e., with the Galilean
transformations. But, for the measured }$N_{m}$\emph{\ the nonrelativistic
theory predicts too small }$N_{s}.$\emph{\ The ''AT relativity'' correctly
predicts the value of }$N_{s}$\emph{\ in both frames but only in the ''e''
coordinatization, while in the ''r'' coordinatization the experimental }$%
N_{s}$\emph{\ and the theoretically predicted }$N_{s}$\emph{\ drastically
differ. The ''TT relativity'' completely agrees with the experiments in all
IFRs and all possible coordinatizations. Thus, only the manifestly covariant
formulation of the special relativity, i.e., the ''TT relativity,'' as the
theory of 4D spacetime with the pseudo-Euclidean geometry, is in a complete
agreement with the experiments.}

\subsection{Another time ''dilatation'' experiments}

The same conclusion can be obtained comparing the other particle lifetime
measurements, e.g., \cite{rossi}, or for the pion lifetime \cite{ayres},
with all three theories. However, as it is already said, all the mentioned
experiments, and not only them but all other too, were designed to test the
''AT relativity.'' Thus in the experiments \cite{rossi}, which preceded to
the experiments \cite{Frisch} and \cite{Easwar}, the relation similar to (%
\ref{radAT1}) is used but with $t_{E}$ replaced by $H_{E}$ (=$vt_{E}$) and $%
\tau _{E}$ (the lifetime of muons in the Earth frame) replaced by $L$ $%
=v\tau _{E}$ ($L$ is the ''average range before decay''), and also the
connection between the lifetimes (\ref{taue}) ($\tau _{E}=\gamma \tau _{\mu
} $) is employed. Obviously the \emph{predictions} of the results in the
experiments \cite{rossi} will depend on the chosen synchronization, since
they deal with the ''AT relativity'' and use the radioactive-decay law in
the form that contains only a part of the distance 4-vector. The \emph{%
predictions }obtained by the use of the ''TT relativity'' will be again
independent on the chosen IFR and the chosen coordinatization. However the
comparison of these experiments \cite{rossi} with the ''TT relativity'' is
difficult since, e.g., they have no data for $t_{\mu }.$ Similarly happens
with the experiments reported in \cite{ayres}.

The lifetime measurements of muons in the g-2 experiments \cite{bailey} are
often quoted as the most convincing evidence for the time dilatation, i.e.,
they are claimed as high-precision evidence for the special relativity.
Namely in the literature the evidence for the time dilatation is commonly
considered as the evidence for the special relativity. The muon lifetime in
flight $\tau $ is determined by fitting the experimental decay electron time
distribution to the six-parameter phenomenological function describing the
normal modulated exponential decay spectrum (their Eq.(1)). Then by the use
of the relation $\tau =\gamma \tau _{0}$ and of $\tau _{0}$ (our $\tau _{\mu
}$), the lifetime at rest (as determined by other workers), they obtained
the time-dilatation factor $\gamma ,$ or the kinematical $\gamma .$ This $%
\gamma $ is compared with the corresponding dynamical $\gamma $ factor ($%
\gamma =(p/m)dp/dE$), which they called $\overline{\gamma }$ (the average $%
\gamma $ value). $\overline{\gamma }$ is determined from the mean rotation
frequency $\overline{f}_{rot}$ by the use of the Lorentz force law (the
''relativistic'' expression); the magnetic field was measured in terms of
the proton NMR frequency $f_{p}$ (for the discussion of $g-2$ experiments
within the traditional ''AT relativity'' see also \cite{newman}). Limits of
order $10^{-3}$ in $(\gamma -\overline{\gamma })/\gamma $ at the kinematical 
$\gamma =29.3$ were set. In that way they also compared the value of the $%
\mu ^{+}$ lifetime at rest $\tau _{0}^{+}$ (from the other precise
measurements) with the value found in their experiment $\tau ^{+}/\overline{%
\gamma },$ and obtained $(\tau _{0}^{+}-\tau ^{+}/\overline{\gamma })/\tau
_{0}^{+}=(2\pm 9)\times 10^{-4},$ (this is the same comparison as the
mentioned comparison of $\gamma $ with $\overline{\gamma }$). They claimed:
''At $95\%$ confidence the fractional difference between $\tau _{0}^{+}$ and 
$\tau ^{+}/\overline{\gamma }$ is in the range $(-1.6-2.0)\times 10^{-3}$.''
and ''To date, this is the most accurate test of relativistic time dilation
using elementary particles.'' The objections to the precision of the
experiments \cite{bailey}, and the remark that a convincing direct test of
special relativity must not assume the validity of special relativity in
advance (in the use of the ''relativistic'' Lorentz force law in the
determination of the mean rotation frequency and thus of $\overline{\gamma }%
, $ and $\tau _{0}$), have been raised in \cite{huang}. The discussion of
these objections is given in \cite{field}.

However, our objections to \cite{bailey} are of a quite different nature.
Firstly, the theoretical relations refer to the ''e'' coordinatization and,
e.g., Eq.(1) in the first paper in \cite{bailey} cannot be transformed in an
appropriate way to the ''r'' coordinatization in order to compare the ''AT
relativity'' in different coordinatizations with the experiments. If only
the exponential factor is considered then this factor is again, as in \cite
{Frisch}, affected by synchrony choice. Although the time $t$ in that
exponential factor may be independent of the chosen synchronization (when $t$
is taken to be the multiple of the mean rotation period $T$), but $\tau $
does not refer to the events that happen at the same spatial point and thus
it is synchrony dependent quantity. This means that in the ''r'' base one
cannot use the relation $\tau =\gamma \tau _{0}$ to find the ''dilatation''
factor $\gamma ,$ but the relation (21) from $\left[ I\right] $ for the time
''dilatation'' in the ''r'' base, $x_{r}^{0}(\tau )=(1+2\beta
_{r})^{1/2}c\tau _{0}$ must be employed. Hence, the whole comparison of $%
\gamma $ with $\overline{\gamma }$ holds only in the ''e'' base; in another
coordinatization the ''AT relativity'' predicts quite different $\tau _{0}$
for the same $x^{0}(\tau ),$ which is inferred from the exponential decay
spectrum.

Let us now examine the measurements \cite{bailey} from the point of view of
the ''TT relativity.'' But for the ''TT relativity'' these experiments are
incomplete and cannot be compared with the theory. Namely, in the ''TT
relativity,'' as already said, it is not possible to find the values of the
muon lifetime in flight $\tau $ by analyses of the measurements of the
radioactive decay distribution, since, there, the radioactive decay law is
written in terms of the spacetime lengths and not with $t$ and $\tau .$
Also, in the ''TT relativity,'' there is not the connection between the muon
lifetime in flight $\tau $ and the lifetime at rest $\tau _{0}$ in the form $%
\tau =\gamma \tau _{0},$ since $\tau ,$ in the ''TT relativity,'' does not
exist as a well defined quantity. Thus, in the ''TT relativity,'' there is
no sense in the use of the relation $\tau =\gamma \tau _{0}$ to determine $%
\gamma .$ An important remark is in place here; in principle, in the ''TT
relativity,'' the same events and the same quantities have to be considered
in different frames of reference, which means that in the muon experiment 
\cite{bailey} the lifetime at rest $\tau _{0}$ refers to the decaying
particle in an accelerated frame and for the theoretical discussion we would
need to use the coordinate transformations connecting an IFR with an
accelerated frame of reference. (An example of the generalized Lorentz
transformation is given in \cite{nelson} but they are written in the ''e''
base and thus not in fully covariant way, i.e., not in the way as we have
written the covariant Lorentz transformation, Eq.(3) in $\left[ I\right] $.)
Furthermore, in the experiments \cite{bailey} the average value of $\gamma $
($\overline{\gamma }$), i.e., the dynamical $\gamma ,$ for the circulating
muons is found by analysis of the bunch structure of the stored muon and the
use of the relation connecting $\overline{\gamma }$ and the mean rotation
frequency $\overline{f}_{rot};$ this relation is obtained by the use of the
expression for the ''relativistic,'' i.e., the ''AT relativity,'' Lorentz
force law, which is expressed by means of the 3-vectors $\mathbf{E}$ and $%
\mathbf{B.}$ However, in contrast to the ''AT relativity,'' and also to the
usual covariant formulation, in the ''TT relativity,'' the covariant Lorentz
force $K^{a}=(q/c)F^{ab}u_{b}$ ($F^{ab}$ is the electromagnetic field tensor
and $u^{b}$ is the 4-velocity of a charge $q;$ all is written in the
abstract index notation, \cite{Wald} and \cite{vanzel}) cannot be expressed
in terms of the 3-vectors $\mathbf{E}$ and $\mathbf{B.}$ Namely, as already
said, in the ''AT relativity'' the real physical meaning is attributed not
to $F^{ab}$ than to the 3-vectors $\mathbf{E}$ and $\mathbf{B,}$ while in
the ''TT relativity'' only covariantly defined quantities do have
well-defined physical meaning both in the theory and in experiments. (The
transformations of the 3-vectors $\mathbf{E}$\ and $\mathbf{B}$ are not
directly connected with the Lorentz transformations of the \emph{whole 4D
tensor quantity} $F^{ab}$ as a geometrical quantity$,$\ but indirectly
through the transformations of \emph{some components} of $F^{ab},$\ and
that, \emph{in the specific coordinatization, the Einstein coordinatization.}
This issue is discussed in \cite{ivezic} and \cite{ivsc98,ivmom}, where it
is also shown that the 3-vector $\mathbf{E}$ ($\mathbf{B}$) in an IFR $S$
and the transformed 3-vector $\mathbf{E}^{\prime }$ ($\mathbf{B}^{\prime }$)
in relatively moving IFR $S^{\prime }$ do not refer to the same physical
quantity in 4D spacetime, i.e., that the conventional transformations of $%
\mathbf{E}$ and $\mathbf{B}$ are the AT.) From \cite{vanzel} and \cite
{ivsc98} one can see how the Lorentz force $K^{a}$ is expressed in terms of
the 4-vectors $E^{a}$ and $B^{a}$ and show when this form corresponds to the
classical expression for the Lorentz force with the 3-vectors $\mathbf{E}$
and $\mathbf{B.}$ Also it can be seen from \cite{ivsc98} that for $B^{\alpha
}\neq 0$ ($B^{\alpha }$ is the representation of $B^{a}$ in the ''e'' base)
it is not possible to obtain $\gamma _{u}=1$ (the 4-velocity of a charge $q$
in the ''e'' base is $u^{\alpha }=(\gamma _{u}c,\gamma _{u}\mathbf{u})$ and $%
\gamma _{u}=(1-u^{2}/c^{2})^{-1/2}$) and the covariant Lorentz force $K^{a}$
can never take the form of the usual magnetic force $\mathbf{F}_{B}.$ Hence
it follows that in the ''TT relativity'' it is not possible to use the
Lorentz force $\mathbf{F}_{B}$ and the usual equation of motion $d(\overline{%
\gamma }m\mathbf{u)/}dt\mathbf{=}q(\mathbf{u}\times \mathbf{B})$ to find the
relation connecting $\overline{\gamma }$ and the mean rotation frequency $%
\overline{f}_{rot},$ and thus to find $\tau _{0}$ from $\tau /\overline{%
\gamma },.$in the way as in \cite{bailey}. The discussion about the
kinematical $\gamma $ (the relation $\tau =\gamma \tau _{0}$) and about the
dynamical $\overline{\gamma }$ (from the use of the Lorentz force) shows
that the measurements \cite{bailey} cannot be compared with the ''TT
relativity.'' But, as we explained before, in contrast to the usual opinion,
these experiments do not confirm the ''AT relativity'' either, since if the
exponential decay spectrum is analyzed in another coordinatization, e.g.,
the ''r'' coordinatization, then, similarly as for the experiments \cite
{Frisch}, one finds that for the given $N_{0}$ the theoretical and the
experimental $N$ differ.

\section{THE MICHELSON-MORLEY EXPERIMENT}

These conclusions will be further supported considering some other
experiments, which, customarily, were assumed to confirm the ''AT
relativity.'' The first one will be the famous Michelson-Morley experiment 
\cite{michels}, and some modern versions of this experiment will be also
discussed.

In the Michelson-Morley experiment two light beams emitted by one source are
sent, by half-silvered mirror $O$, in orthogonal directions. These partial
beams of light traverse the two equal (of the length $L$) and perpendicular
arms $OM_{1}$ (perpendicular to the motion) and $OM_{2}$ (in the line of
motion) of Michelson's inteferometer and the behaviour of the interference
fringes produced on bringing together these two beams after reflection on
the mirrors $M_{1}$ and $M_{2}$ is examined. In order to avoid the influence
of the effect that the two lengths of arms are not exactly equal the entire
inteferometer is rotated through $90^{0}.$ Then any small difference in
length becomes unimportant. The experiment consists of looking for a shift
of the intereference fringes as the apparatus is rotated. The expected
maximum shift in the number of fringes (the measured quantity) on a $90^{0}$%
\ rotation is 
\begin{equation}
\bigtriangleup N=\bigtriangleup (\phi _{2}-\phi _{1})/2\pi ,  \label{delfi}
\end{equation}
where $\bigtriangleup (\phi _{2}-\phi _{1})$ is the change in the phase
difference when the interferometer is rotated through $90^{0}.$ $\phi _{1}$
and $\phi _{2}$ are the phases of waves moving along the paths $OM_{1}O$ and 
$OM_{2}O,$ respectively.

\subsection{The nonrelativistic approach}

In the nonrelativistic approach the speed of light in the preferred frame is 
$c.$ Then, on the ether hypothesis, the speed of light, in the Earth frame,
i.e.,\textbf{\ }in the rest frame of the interferometer (the $S$\ frame), on
the path along an arm of the Michelson interferometer oriented perpendicular
to its motion at velocity $\mathbf{v}$ relative to the preferred frame (the
ether) is $(c^{2}-v^{2})^{1/2};$ the Earth together with the inteferometer
moving with velocity $\mathbf{v}$ through the ether is equivalent to the
inteferometer at rest with the ether streaming through it with velocity $-%
\mathbf{v.}$ Since in $S$ both waves are brought together to the same
spatial point the phase difference $\phi _{2}-\phi _{1}$ is determined only
by the time difference $t_{2}-t_{1};$ $\phi _{2}-\phi _{1}=2\pi
(t_{2}-t_{1})/T,$ where $t_{1}$ and $t_{2}$ are the times required for the
complete trips $OM_{1}O$ and $OM_{2}O,$ respectively, and $T($=$\lambda /c)$
is the period of vibration of the light. From the known speed of light one
finds that $t_{1}$ is 
\begin{equation}
t_{1}=2L/c(1-v^{2}/c^{2})^{1/2}.  \label{tejedan}
\end{equation}
Similarly, the speed of light on the path $OM_{2}$ is $c-v,$ and on the
return path is $c+v,$ giving that 
\begin{equation}
t_{2}=2L/c(1-v^{2}/c^{2}).  \label{tedva}
\end{equation}
We see that according to the nonrelativistic approach the time $t_{1}$ is a
little less than the time $t_{2},$ even though the mirrors $M_{1}$ and $%
M_{2} $ are equidistant from $O.$ To order $v^{2}/c^{2}$ the difference in
the times is $t_{2}-t_{1}=(L/c)(v^{2}/c^{2}).$ Inserting it into $%
\bigtriangleup N$ (\ref{delfi}) (the measured quantity $\bigtriangleup N,$
when the phase difference is determined by the time difference, is $%
\bigtriangleup N=2(t_{2}-t_{1})c/\lambda $) we find, to the same order $%
v^{2}/c^{2}$, that 
\begin{equation}
\bigtriangleup N\simeq (2L/\lambda )(v^{2}/c^{2}).  \label{delclas}
\end{equation}
This result is obtained by the classical analysis in the Earth frame (the
interferometer rest frame).

Let us now consider the same experiment in the preferred frame (the $%
S^{\prime }$\ frame).Since in the nonrelativistic theory the two frames are
connected by the Galilean transformations, it follows that the corresponding
times in both frames are equal, $t_{1}=t_{1}^{\prime }$ and $%
t_{2}=t_{2}^{\prime },$ whence $t_{2}-t_{1}=t_{2}^{\prime }-t_{1}^{\prime }$
and, supposing that again the phase difference is determined only by the
time difference, $\bigtriangleup N=\bigtriangleup N^{\prime }.$ However, for
the further purposes, it is worth to find explicitly $t_{1}^{\prime }$ and $%
t_{2}^{\prime }$ considering the experiment directly in the preferred frame.
Since the speed of light in the preferred frame is $c,$ the preferred-frame
observer considers that the light travels a distance $ct_{1}^{\prime }/2$\
along the hypotenuse of a triangle; in the same time $t_{1}^{\prime }/2$ the
mirror $M_{1}$ moves to $M_{1}^{\prime }$, i.e., to the right a distance $%
vt_{1}^{\prime }/2,$ and from the right triangle this observer finds $%
t_{1}^{\prime }/2=L/c(1-v^{2}/c^{2})^{1/2}.$ The return trip is again along
the hypotenuse of a triangle\emph{\ }and the return time is\emph{\ }again $%
=t_{1}^{\prime }/2.$ The total time for such a zigzag path is, as it must
be, $t_{1}^{\prime }=$ $t_{1}$ (\ref{tejedan}) (the half-silvered mirror $O$
moved to $O^{\prime }$ in $t_{1}^{\prime }$). For the arm oriented parallel
to its motion the preferred-frame observer considers that the light, when
going from $O$ to $M_{2}^{\prime }$, must traverse a distance $%
L+vt_{3}^{\prime }$ at the speed $c,$ whence $L+vt_{3}^{\prime
}=ct_{3}^{\prime }$ and $t_{3}^{\prime }=L/(c-v)$. In a like manner, the
time $t_{4}^{\prime }$ for the return trip is $t_{4}^{\prime }=L/(c+v).$ The
total time $t_{2}^{\prime }=t_{3}^{\prime }+t_{4}^{\prime }$ is, as it must
be, equal to $t_{2}$ (\ref{tedva}) (the half-silvered mirror $O$ moved to $%
O^{\prime \prime }$ in $t_{2}^{\prime }$). This discussion shows that the
nonrelativistic theory is a consistent theory giving the same $%
\bigtriangleup N$ in both frames. However it does not agree with the
experiment. Namely Michelson and Morley found from their experiment that was
no observable fringe shift.

From the theoretical point of view it is interesting to mention an analysis
of the Michelson-Morley experiment which is given in \cite{schum}. There,
the paths of light are examined in the case when the experiment is viewed
from a frame in which the apparatus is moving at velocity $v$ (our $%
S^{\prime }$ frame). It is inquired whether the half-silvered mirror $O$
correctly reflects the light to and from the interferometer arms, such that
light travels in the appropriate ''triangular path'' in the transverse arm,
and correctly brings the longitudinal ray into line with the transverse ray
at the detector. The result is that if in the classical analysis the
half-silvered mirror is set to exactly $45^{0},$ then the transverse ray
will ''overshoot'' the desired trajectory while the longitudinal ray will
''undershoot.'' The interference pattern will be dependent on the position
of the detector since there is a divergence of the interfering light rays.
The ray angles on the way to the detector are given by Eq.(16) in \cite
{schum} The difference in these angles is exceedingly small (second order in 
$v/c$), and hence negligible in the usual Michelson-Morley experiment.

At this point it has to be noted that there are more serious objections to
the traditional derivation of $\bigtriangleup N$ in the nonrelativistic
theory and in $S^{\prime }$ than the one mentioned by \cite{schum}. These
objections are usually overlooked and we only briefly sketch them here.
Firstly, in $S^{\prime }$ the waves are not brought together to the same
spatial point and consequently the phase difference is not determined only
by the time difference. Strictly speaking the increment of phase $\phi
_{1}^{\prime }$ for the trip $OM_{1}^{\prime }O^{\prime }$ is $\phi
_{1}^{\prime }=(\omega _{OM_{1}^{\prime }}^{\prime }t_{1}^{\prime }/2-%
\mathbf{k}_{OM_{1}^{\prime }}^{\prime }\mathbf{l}_{OM_{1}^{\prime }}^{\prime
})+(\omega _{M_{1}^{\prime }O^{\prime }}^{\prime }t_{1}^{\prime }/2-\mathbf{k%
}_{M_{1}^{\prime }O^{\prime }}^{\prime }\mathbf{l}_{M_{1}^{\prime }O^{\prime
}}^{\prime }),$ and similarly the increment of phase $\phi _{2}^{\prime }$
for the trip $OM_{2}^{\prime }O^{\prime \prime }$ is $\phi _{2}^{\prime
}=(\omega _{OM_{2}^{\prime }}^{\prime }t_{3}^{\prime }-\mathbf{k}%
_{OM_{2}^{\prime }}^{\prime }\mathbf{l}_{OM_{2}^{\prime }}^{\prime
})+(\omega _{M_{2}^{\prime }O^{\prime \prime }}^{\prime }t_{4}^{\prime }-%
\mathbf{k}_{M_{2}^{\prime }O^{\prime \prime }}^{\prime }\mathbf{l}%
_{M_{2}^{\prime }O^{\prime \prime }}^{\prime }),$ where $\omega
_{OM_{1}^{\prime }}^{\prime },$ $\mathbf{k}_{OM_{1}^{\prime }}^{\prime },$
and $\mathbf{l}_{OM_{1}^{\prime }}^{\prime }$ are the angular frequency, the
wave 3-vector and the distance 3-vector ($\overrightarrow{OM_{1}^{\prime }}$%
), respectively, of the wave on the trip $OM_{1}^{\prime },$ etc.. What is
overlooked in the usual derivation of $\bigtriangleup N$ in the
nonrelativistic theory is that not all $\omega ^{\prime }$ are the same due
to the classical Doppler effect of inertial motion of a source and of a
mirror in the $S^{\prime }$ frame, and that the classical aberration of
light has to be taken into account when different $\mathbf{k}^{\prime }$ in $%
S^{\prime }$ are determined (this is, in fact, considered in \cite{schum}).
We shall not examine the mentioned changes of the classical derivation since 
$\bigtriangleup N,$ obtained with these changes, will be again different
from zero.

It is possible to look at the Michelson-Morley experiment from another point
of view; the light ray going both ways in one of the arms of the
interferometer can be considered as a clock, a light clock, with the period
determined by the return time of the light ray. The experiment is then
considered as the comparison of the frequencies of two clocks, and it shows
that the relative frequency does not change by a rotation of the
interferometer. Such point of view is important for the interpretation of
the modern versions of the Michelson-Morley experiment.

\subsection{The ''AT relativity'' approach}

Next we examine the same experiment from the ''AT relativity'' viewpoint.
Again, as in the discussion of the ''muon'' experiment, we consider this
experiment in both frames and in both coordinatizations as well. First the
''e'' coordinatization in both frames will be explored. It has to be noted
that the experiment is usually discussed only in the ''e'' base, and again,
as in the nonrelativistic theory, the phase difference $\phi _{2}-\phi _{1}$
is considered to be determined only by the time difference $t_{2}-t_{1}.$
Further, in contrast to the nonrelativistic theory, in the ''AT relativity''
and in the ''e'' base it is postulated (Einstein's second postulate) that
light \emph{always} travels with speed $c.$

Hence in the $S$\ frame (the rest frame of the interferometer)\textbf{\ }$%
t_{1}=2L/c$ and $t_{2}=2L/c,$ and, with the assumption that only the time
difference $t_{2}-t_{1}$ matters, it follows that $\bigtriangleup N$=$0,$ in
agreement with the experiment. In the $S^{\prime }$\ frame (the preferred
frame)\textbf{\ }the time $t_{1}^{\prime }$ is determined in the same way as
in the nonrelativistic theory, i.e., supposing that a zigzag path is taken
by the light beam in a moving ''light clock''. Thus, the light-travel time $%
t_{1}^{\prime }$ is exactly equal to that one in the nonrelativistic theory, 
$t_{1}^{\prime }=2L/c(1-v^{2}/c^{2})^{1/2}.$ Comparing with \textbf{\ }$%
t_{1}=2L/c$ we see that, in contrast to the nonrelativistic theory, it takes
a longer time for light to go from end to end in the moving clock than in
the stationary clock, $t_{1}^{\prime }=t_{1}/(1-v^{2}/c^{2})^{1/2}=\gamma
t_{1},$ see, e.g., \cite{Feyn} p.15-6, \cite{Kittel} p.359, or an often
cited paper on modern tests of special relativity \cite{haugan}. This is the
usual way in which it is shown how, in the ''AT relativity,'' the time
dilatation is forced upon us by the constancy of the speed of light.
However, in the ''AT relativity,'' the light-travel time $t_{2}^{\prime }$
is determined by invoking the Lorentz contraction; it is argued that a
preferred frame observer measures the length of the arm oriented parallel to
its motion to be contracted to a length $L^{\prime }=L(1-v^{2}/c^{2})^{1/2},$
see, e.g., \cite{haugan}$.$ Then $t_{2}^{\prime }$ is determined in the same
way as in the nonrelativistic theory but with $L^{\prime }$ replacing the
rest length $L,$ $t_{2}^{\prime }=(L^{\prime }/(c-v))+(L^{\prime
}/(c+v))=2L/c(1-v^{2}/c^{2})^{1/2}=t_{1}^{\prime },$ whence $t_{2}^{\prime
}-t_{1}^{\prime }=0$ and $\bigtriangleup N^{\prime }$=$0,$ as in the $S$
frame. We quoted such usual derivation in order to illustrate how the time
dilatation and the Lorentz contraction are used in the ''AT relativity'' to
show the agreement between the theory and the famous Michelson-Morley
experiment. Although this procedure is generally accepted by the majority of
physicists as the correct one and quoted in all textbooks on the subject, we
note that such an explanation of the null result of the experiment is very
awkward and does not use at all the 4D symmetry of the spacetime; the
derivation deals with the temporal and spatial distances as well defined
quantities, i.e., in a similar way as in the prerelativistic physics, and
then in an artificial way introduces the changes in these distances due to
the motion.

To better illustrate the preceding assertions we derive the same results for 
$t_{1}^{\prime },$ $t_{2}^{\prime }$ and $\bigtriangleup N^{\prime }$ in
another way too. It starts with 4D quantities, but then, as shown in $\left[
I\right] $ Sec.4.2, connects only some parts of the distance 4-vectors,
i.e., the temporal distances, in two relatively moving frames using Eq.(20)
from $\left[ I\right] $ for the time dilatation in the ''e'' base instead of
the complete Lorentz transformation. Let now $A,$ $B$ and $A_{1}$ denote the
events; the departure of the transverse ray from the half-silvered mirror $%
O, $ the reflection of this ray on the mirror $M_{1}$ and the arrival of
this beam of light after the round trip on the half-silvered mirror $O,$
respectively. In the same way we have, for the longitudinal arm of the
inteferometer, the corresponding events $A,$ $C$ and $A_{2}.$ Then, from
Eq.(20) in $\left[ I\right] $, one finds $t_{1}^{\prime }=\gamma t_{1},$ $%
t_{2}^{\prime }=\gamma t_{2}=t_{1}^{\prime }$ and consequently $%
\bigtriangleup N^{\prime }$=$0.$ But, note, that in both mentioned
derivations in the ''AT relativity,'' the fequencies of the waves are
supposed to be the same in $S$ and $S^{\prime },$ i.e., the Doppler effect
is not taken into account, and the contributions of $\mathbf{k}^{\prime }$
and $\mathbf{l}^{\prime }$ to the increments of phase in $S^{\prime }$ are
neglected, i.e., the consideration of the aberration of light in the
determination of different $\mathbf{k}^{\prime }$ in $S^{\prime }$ is not
performed. Obviously, in the ''AT relativity,'' the same procedure is
applied to the calculation of $\phi _{1}^{\prime },$ $\phi _{2}^{\prime }$
and $\bigtriangleup N^{\prime }$ in $S^{\prime }$ as in the nonrelativistic
theory; only, in an artificial way, the Lorentz contraction and the time
dilatation are introduced into the calculation.

The same experiment can be examined in the ''r'' coordinatization\textbf{. }%
This synchronization\textbf{\ }is an asymmetric synchronization, which leads
to an asymmetry in the measured one-way speed of light, but the average
speed of light on any round trip is independent of the synchronization
procedure employed, and is $=c.$ In the Michelson-Morley experiment the
measured phase difference between the phases on the round trips $OM_{1}O$
and $OM_{2}O$ in $S,$ the rest frame of the interferometer, is synchrony
independent, since both waves are brought together to the same spatial
point. Hence, one concludes that the same result $\bigtriangleup N$=$0$ will
be obtained in the $S$ frame in the ''r'' base as in the ''e''
coordinatization. However in the $S^{\prime }$ frame such independence on
the used coordinatization cannot be expected. We shall not discuss this
issue here for the sake of saving space, and since there are some more
important problems in the traditional ''AT relativity'' derivation of $%
\bigtriangleup N$.

As already mentioned, it is shown in \cite{schum} that the classical
analysis of the interference, in the frame in which the apparatus is in
motion, predicts a divergence of the interfering light rays on the way to
the detector. In contrast to this result, the exact parallelness of the
longitudinal and the transverse rays is obtained in \cite{schum}, but only
in the case when, in addition to the usual analysis in the ''AT
relativity,'' the Lorentz contraction ''tilt'' of the moving half-silvered
mirror is taken into account. The analysis in \cite{schum} actually takes
into account the aberration of light, which is overlooked in the traditional
derivation in the ''AT relativity'' in the same way as it is overlooked in
the nonrelativistic theory. However this analysis is performed in the ''e''
coordinatization and the Lorentz contraction ''tilt'' of the moving
half-silvered mirror, that is required for the exact parallelness of the
rays, is considered in that coordinatization. In another coordinatization,
e.g., in the ''r'' coordinatization, the Lorentz ''contraction'' of the
moving half-silvered mirror will be different, it can become a
''dilatation,'' and one can expect a divergence of the interfering light
rays on the way to the detector. But, as it is already said, the effect is
exceedingly small (second order in $v/c$), and hence negligible in the usual
Michelson-Morley experiment, and therefore it will not be discussed in more
detail.

\subsubsection{Driscoll's non-null fringe shift}

In \cite{drisc} the usual ''AT relativity'' calculation in the ''e'' base
(see the discussion above and also \cite{Feyn}, \cite{Kittel}, \cite{haugan}%
) of the fring shift in the Michelson-Morley experiment is repeated, and, of
course, the observed null fringe shift independent of changes of $v$, the
relative velocity of $S$ and $S^{\prime },$ and/or $\theta ,$ the angle that
the undivided ray from the source to the beam divider makes with $\mathbf{v,}
$ is obtained. However, it is noticed in \cite{drisc} that in such a
traditional calculation of $\bigtriangleup (\phi _{2}-\phi _{1})$ (the
change in the phase difference when the interferometer is rotated through $%
90^{0}$) only path lengths (optical or geometrical), i.e., the temporal
distances, are considered, while the Doppler effect on wavelength in the $%
S^{\prime }$ frame, in which the interferometer is moving, is not taken into
account. Then the same calculation of $\bigtriangleup (\phi _{2}^{\prime
}-\phi _{1}^{\prime }),$ as the traditional one, is performed in \cite{drisc}%
, but determing the increment of phase along some path, e.g. $OM_{1}^{\prime
}$, not only by the segment of geometric path length (i.e., the temporal
distance for that path) than also by the wavelength in that segment (i.e.,
the frequency of the wave in that segment). Accordingly the phase difference
(in our notation) $\phi _{1}^{\prime }-\phi _{2}^{\prime },$ in the $%
S^{\prime }$ frame, between the ray along the vertical path $OM_{1}^{\prime
}O^{\prime },$ and that one along the longitudinal path $OM_{2}^{\prime
}O^{\prime \prime },$ respectively, is found (see \cite{drisc}) to be 
\begin{equation}
(\phi _{1}^{\prime }-\phi _{2}^{\prime })_{(b)}/2\pi =2(L\nu
/c)(1+\varepsilon +\beta ^{2})-2(L\nu /c)(1+2\beta ^{2})=2(L\nu
/c)(\varepsilon -\beta ^{2}),  \label{drisbe}
\end{equation}
Eqs.(23-25) in \cite{drisc}, where $L$ is the length of the segment $OM_{2}$
and $\overline{L}=L(1+\varepsilon )$ is the length of the arm $OM_{1}$ ($L,$ 
$\overline{L}$ and $\nu $ are determined in the rest frame of the
interferometer). In this expression the Doppler effect of $\mathbf{v}$ on
the frequencies, and the Lorentz contraction of the longitudinal arm, are
taken into account. In a like manner Driscoll finds the phase difference in
the case when the interferometer is rotated through $90^{0}$ 
\begin{equation}
(\phi _{1}^{\prime }-\phi _{2}^{\prime })_{(a)}/2\pi =2(L\nu
/c)(1+\varepsilon +2\beta ^{2})-2(L\nu /c)(1+\beta ^{2})=2(L\nu
/c)(\varepsilon +\beta ^{2}),  \label{drisca}
\end{equation}
Eqs.(19-21) in \cite{drisc}. Hence, it is found in \cite{drisc} a
''surprising'' non-null fringe shift 
\begin{equation}
\bigtriangleup N^{\prime }=\bigtriangleup (\phi _{2}^{\prime }-\phi
_{1}^{\prime })/2\pi =4(L\nu /c)\beta ^{2},  \label{shift}
\end{equation}
where $\bigtriangleup (\phi _{2}^{\prime }-\phi _{1}^{\prime })=(\phi
_{1}^{\prime }-\phi _{2}^{\prime })_{(b)}-(\phi _{1}^{\prime }-\phi
_{2}^{\prime })_{(a)},$ and we see that the entire fringe shift is due to
the Doppler shift. From the non-null result (\ref{shift}) the author of \cite
{drisc} concluded: ''that the Maxwell-Einstein electromagnetic equations and
special relativity jointly are disproved, not confirmed, by the
Michelson-Morley experiment.'' However such a conclusion cannot be drawn
from the result (\ref{shift}). The origin of the appearance of $%
\bigtriangleup N^{\prime }\neq 0$ (\ref{shift}) is quite different than that
considered in \cite{drisc}, and it will be explained below. Note that in 
\cite{schum} the changes in the usual derivation of $\bigtriangleup
N^{\prime },$ which are caused by the aberration of light, are considered,
while \cite{drisc} investigates those changes which are caused by the
Doppler effect. Both changes are examined only in the ''e'' base, and both
would be different in, e.g., the ''r'' base. This means that $\bigtriangleup
N^{\prime }$ in $S^{\prime }$ will be dependent on the chosen
synchronization, and consequently that the ''AT relativity'' is not capable
to explain in a satisfactory manner the results of the Michelson-Morley
experiment.

\subsection{The ''TT relativity'' approach}

Next we examine the Michelson-Morley experiment from the ''TT relativity''
viewpoint. Then the relevant quantity is the phase of a light wave, and it
is (when written in the abstract index notation, see $\left[ I\right] $ and 
\cite{Wald}) 
\begin{equation}
\phi =k^{a}g_{ab}l^{b},  \label{phase}
\end{equation}
where $k^{a}$ is the propagation 4-vector, $g_{ab}$ is the metric tensor and 
$l^{b}$ is the distance 4-vector. (We note that in the ''TT relativity'' the
light waves are described by the 4-vectors $E^{a}(x^{b})$ and $B^{a}(x^{b})$
of the electric and magnetic fields $(E^{a}(x^{b})=E_{0}^{a}\exp
(ik^{b}x_{b})),$ while the ''AT relativity'' works with the 3-vectors $%
\mathbf{E(r,}t)$ and $\mathbf{B(r,}t)$ $(\mathbf{E(r,}t)=\mathbf{E}_{0}\exp
(i(\mathbf{kr-}\omega t))),$ as in the prerelativistic physics.) The
traditional derivation of $\bigtriangleup N$ (in the ''AT relativity'')
deals only with the calculation of $t_{1}$ and $t_{2}$ in $S$ and $S^{\prime
},$ but does not take into account either the changes in frequencies due to
the Doppler effect or the aberration of light. The ''AT relativity''
calculations in \cite{drisc} and \cite{schum} improve the traditional
procedure taking into account the changes in frequencies \cite{drisc}, and
the aberration of light \cite{schum}, but only in the ''e'' base. None of
the ''AT relativity'' calculations deals with the covariantly defined 4D
quantities, in this case, the covariantly defined phase (\ref{phase}), and
it will be shown here that the non-null theoretical result obtained in \cite
{drisc} is a consequence of that fact. In the ''TT relativity'' neither the
Doppler effect nor the aberration of light exist separately as well defined
physical phenomena. The separate contributions to $\phi $ of the $\omega t$
factors and $\mathbf{kl}$ factors are, in general case, meaningless in the
''TT relativity'' and only their indivisible unity, the phase (\ref{phase}),
is meaningful quantity; it is invariant upon the covariant 4D Lorentz
transformations (Eq.(3) in $\left[ I\right] $ and Sec.3 here)$,$ and, as it
is written in the abstract index notation, the phase (\ref{phase}) does not
depend on the chosen coordinatization in the considered IFR. (All quantities
in (\ref{phase}), i.e., $k^{a}$, $g_{ab}$ and $l^{b},$ are the 4D tensor
quantities that correctly transform upon the covariant 4D Lorentz
transformations (Eq.(3) in $\left[ I\right] $ and Sec.3 here)$,$ which means
that in all relatively moving IFRs always \emph{the same 4D quantity}, e.g., 
$k^{a},$ or $l^{b},$ is considered. This is not the case in the ''AT
relativity'' where, for example, the relation $t_{1}^{\prime }=\gamma t_{1}$
is not the Lorentz transformation of some 4D quantity, and $t_{1}^{\prime }$%
\emph{\ and }$t_{1}$\emph{\ do not correspond to the same 4D quantity}
considered in $S^{\prime }$ and $S$ respectively.) Only in the ''e''
coordinatization the $\omega t$ and $\mathbf{kl}$ factors can be considered
separately. Therefore, and in order to retain the similarity with the
prerelativistic and the ''AT relativity'' considerations, we first determine 
$\phi $ (\ref{phase}) in the ''e'' base and in the $S$ frame (the rest frame
of the interferometer). Then $k_{ABe}^{\mu }$ and $l_{ABe}^{\mu }$ (the
representations of $k_{AB}^{a}$ and $l_{AB}^{a}$ in the ''e'' base and in $S$%
) for the wave on the trip $OM_{1}$ ( $A$ and $B$ are the corresponding
events for that trip, as mentioned previously) $k_{ABe}^{\mu }=(\omega
/c,0,2\pi /\lambda ,0)$ and $l_{ABe}^{\mu }=(ct_{M_{1}},0,\overline{L},0),$
while for the wave on the return trip $M_{1}O,$ (the events $B$ and $A_{1}$) 
$k_{BA_{1}e}^{\mu }=(\omega /c,0,-2\pi /\lambda ,0)$ and $l_{BA_{1}e}^{\mu
}=(ct_{M_{1}},0,-\overline{L},0)$). Hence the increment of phase $\phi _{1e}$%
, for the the round trip $OM_{1}O,$ is 
\[
\phi _{1e}=k_{ABe}^{\mu }l_{\mu ABe}+k_{BA_{1}e}^{\mu }l_{\mu
BA_{1}e}=2(-\omega t_{M_{1}}+(2\pi /\lambda )\overline{L}), 
\]
where $\omega $ is the angular frequency and, for the sake of comparison
with \cite{drisc}, the length of the arm $OM_{1}$ is taken to be $\overline{L%
}=L(1+\varepsilon ),$ and $L$ is the length of the segment $OM_{2}.$ In a
like manner we find $k_{ACe}^{\mu }$ and $l_{ACe}^{\mu }$ for the wave on
the trip $OM_{2},$ (the corresponding events for the round trip $OM_{2}O$
are $A,$ $C$ and $A_{2}$) as $k_{ACe}^{\mu }=(\omega /c,2\pi /\lambda ,0,0)$
and $l_{ACe}^{\mu }=(ct_{M_{2}},L,0,0),$ while for the wave on the return
trip $M_{2}O,$ $k_{CA_{2}e}^{\mu }=(\omega /c,-2\pi /\lambda ,0,0)$ and $%
l_{CA_{2}e}^{\mu }=(ct_{M_{2}},-L,0,0)$), whence 
\[
\phi _{2e}=k_{ACe}^{\mu }l_{\mu ACe}+k_{CA_{2}e}^{\mu }l_{\mu
CA_{2}e}=2(-\omega t_{M_{2}}+(2\pi /\lambda )L), 
\]
and thus 
\begin{equation}
\phi _{1e}-\phi _{2e}=-2\omega (t_{M_{1}}-t_{M_{2}})+2(2\pi /\lambda )(%
\overline{L}-L).  \label{phasdif}
\end{equation}
Particularly for $\overline{L}=L,$ and consequently $t_{M_{1}}=t_{M_{2}},$
one finds $\phi _{1e}-\phi _{2e}=0.$ It can be easily shown that the same
difference of phase (\ref{phasdif}) is obtained in the case when the
interferometer is rotated through $90^{0},$ whence we find that $%
\bigtriangleup (\phi _{1e}-\phi _{2e})$=$0,$ and $\bigtriangleup N_{e}=0.$
Since, according to the construction, $\phi $ (\ref{phase}) is a Lorentz
scalar, and does not depend on the chosen coordinatization in a considered
IFR, we immediately conclude, without calculation, that 
\begin{equation}
\bigtriangleup N_{e}^{\prime }=\bigtriangleup N_{r}=\bigtriangleup
N_{r}^{\prime }=\bigtriangleup N_{e}=0,  \label{delshif}
\end{equation}
which is in a complete agreement with the Michelson-Morley experiment.

\subsubsection{Explanation of Driscoll's non-null fringe shift and of the
null fringe shift obtained in the conventional ''AT relativity'' calculation}

Driscoll's improvement of the traditional ''AT relativity'' derivation of
the fringe shift can be easily obtained from our covariant approach taking
only the product $k_{e}^{\prime 0}l_{0e}^{\prime }$ in the calculation of
the increment of phase $\phi _{e}^{\prime }$ in $S^{\prime }$ in which the
apparatus is moving. Thus in $S^{\prime }$ $k_{ABe}^{\prime \mu }=(\gamma
\omega /c,-\beta \gamma \omega /c,2\pi /\lambda ,0)$ and $l_{ABe}^{\prime
\mu }=(\gamma ct_{M_{1}},-\beta \gamma ct_{M_{1}},\overline{L},0),$ and also 
$k_{BA_{1}e}^{\prime \mu }=(\gamma \omega /c,-\beta \gamma \omega /c,-2\pi
/\lambda ,0)$ and $l_{BA_{1}e}^{\prime \mu }=(\gamma ct_{M_{1}},-\beta
\gamma ct_{M_{1}},-\overline{L},0),$ giving that 
\begin{equation}
(-1/2\pi )(k_{ABe}^{\prime 0}l_{0ABe}^{\prime }+k_{BA_{1}e}^{\prime
0}l_{0BA_{1}e}^{\prime })=2\gamma ^{2}\nu t_{M_{1}}\simeq 2(L\nu
/c)(1+\varepsilon +\beta ^{2}),  \label{truedri}
\end{equation}
which is exactly equal to Driscoll's result $\bigtriangleup P_{Hb},$ for our
notation see (\ref{drisbe}). In a like manner one finds that 
\begin{equation}
(-1/2\pi )(k_{ACe}^{\prime 0}l_{0ACe}^{\prime }+k_{CA_{2}e}^{\prime
0}l_{0CA_{2}e}^{\prime })=2\gamma ^{2}(\nu t_{M_{2}}+\beta ^{2}L/\lambda
)\simeq 2(L\nu /c)(1+2\beta ^{2}),  \label{trepar}
\end{equation}
which is Driscoll's result $\bigtriangleup P_{\Xi b},$ see (\ref{drisbe}).
In the same way we can find in $S^{\prime }$ Driscoll's result (\ref{drisca}%
) and finally the non-null fringe shift, Eq.(\ref{shift}). The same
calculation of $k_{e}^{\prime i}l_{ie}^{\prime },$ the contribution of the
spatial parts of $k_{e}^{\prime \mu }$ and $l_{\mu e}^{\prime }$ to $%
\bigtriangleup N_{e}^{\prime },$ shows that this term exactly cancel the $%
k_{e}^{\prime 0}l_{0e}^{\prime }$ contribution (Driscoll's non-null fringe
shift (\ref{shift})), yielding that $\bigtriangleup N_{e}^{\prime }=0.$ We
note that the calculation in \cite{drisc} actually assumes that $%
k_{e}^{0}l_{0e}$ and $k_{e}^{\prime 0}l_{0e}^{\prime }$ refer to the same
quantity measured by the observers in $S$ and $S^{\prime }$; of course, it
is supposed that $S$ and $S^{\prime }$ are connected by the Lorentz
transformation, and consequently that the quantities $k_{e}^{0}l_{0e}$ and $%
k_{e}^{\prime 0}l_{0e}^{\prime }$ are connected by the Lorentz
transformation as well. However the relations (\ref{truedri}) and (\ref
{trepar}) are not the Lorentz transformation of some 4D quantity, and really 
$k_{e}^{0}l_{0e}$ \emph{and} $k_{e}^{\prime 0}l_{0e}^{\prime }$ \emph{do not
refer to the same 4D quantity} considered in $S$ and $S^{\prime }$
respectively. Thus in this case too the quantities $k_{e}^{0}l_{0e}$ and $%
k_{e}^{\prime 0}l_{0e}^{\prime }$ are connected by the AT, and, as Gamba 
\cite{gamba} says, whenever two quantities, which are connected by the AT,
are considered to refer to the same physical quantity in 4D spacetime we
have \emph{the case of mistaken identity.}

The traditional ''AT relativity'' analysis of the experiment deals only with
the calculation of $t_{1}$ and $t_{2}$ in $S$ and $S^{\prime },$ ($t_{1}$ in 
$S$ is $=2t_{M_{1}}$ and $ct_{M_{1}}$ is only zeroth component $l_{ABe}^{0}$
($=ct_{M_{1}}$) of $l_{ABe}^{\mu },$ and similarly for $t_{2}$ and $%
ct_{M_{2}}$). Furthermore, such traditional ''AT relativity'' calculation
simply connects, e.g., zeroth component $l_{ABe}^{0}$ in $S$ and $%
l_{ABe}^{\prime 0}$ in $S^{\prime }$, or $l_{ACe}^{0}$ and $l_{ACe}^{\prime
0},$ etc., by the relation for the time ''dilatation'' in the ''e'' base, $%
l_{ABe}^{\prime 0}=\gamma ct_{M_{1}}$, and also $l_{ACe}^{\prime 0}=\gamma
ct_{M_{2}}$, although $l_{ACe}^{\prime 0},$ when determined by the Lorentz
transformation, is $l_{ACe}^{\prime 0}=\gamma (ct_{M_{2}}-\beta L),$ which
then yields that $t_{1}^{\prime }=\gamma t_{1},$ $t_{2}^{\prime }=\gamma
t_{2}=t_{1}^{\prime },$ and consequently one finds the null fringe shift $%
\bigtriangleup N^{\prime }$=$0.$ It is clear from this discussion that, in
contrast to the usual opinion, the quantities, e.g., $t_{1}^{\prime }$ and $%
t_{1},$ etc., do not refer to the same quantity, which is measured in
relatively moving IFRs $S^{\prime }$ and $S,$ but they are different 4D
quantities that are not connected by the Lorentz transformation. Therefore
the agreement of he traditional ''AT relativity'' calculation with the
results of the Michelson-Morley experiment is not the ''true'' agreement,
but an ''apparent'' agreement that is achieved by an incorrect procedure.

\emph{The whole discussion about the the Michelson-Morley experiment reveals
that, contrary to the generaly accepted opinion, the Michelson-Morley
experiment does not confirm the validity of the traditional Einstein
approach to the special relativity, i.e., the ''AT relativity,'' than it
confirms the validity of a compete covariant approach, both in the theory
and experiments, i.e., it confirms the ''TT relavitity.'' }

\subsection{The modern laser versions}

The modern laser versions of the Michelson-Morley experiment, e.g., \cite
{jaseja} and \cite{brillet}, are always interpreted according to the ''AT
relativity.'' They rely on highly monochromatic (maser) laser frequency
metrology rather than optical interferometry; the measured quantity is not
the maximum shift in the number of fringes than a beat frequency variation
and the associated (maser) laser-frequency shift. In \cite{jaseja} the
authors recorded the variations in beat frequency between two optical maser
oscillators when rotated through $90^{0}$ in space; the two maser cavities
are placed orthogonally on a rotating table and they can be considered as
two light clocks. It is stated in \cite{jaseja} that the highly
monochromatic frequencies of masers; ''...allow very sensitive detection of
any change in the round-trip optical distance between two reflecting
surfaces.'' and that the comparison of the frequencies of two masers allows:
''...a very precise examination of the isotropy of space with respect to
light propagation.'' The result of this experiment was: ''... there was no
relative variation in the maser frequencies associated with orientation of
the earth in space greater than about 3 kc/sec.'' Similarly \cite{brillet}
compares the frequencies of a He-Ne laser locked to the resonant frequency
of a higly stable Fabry-Perot cavity (the meter-stick, i.e., ''etalon of
length'') and of a $CH_{4}$ stabilized ''telescope-laser'' frequency
reference system. The beat frequency of the isolation laser ($CH_{4}$
stabilized-laser) with the cavity-stabilized laser was the measured
quantity; a beat frequency variation is considered when the direction of the
cavity length is rotated. The authors of \cite{brillet}, in the same way as 
\cite{jaseja}, consider their experiment as: ''isotropy of space
experiment.'' Namely it is stated in \cite{brillet} that: ''Rotation of the
entire electro-optical system maps any cosmic directional anisotropy of
space into a corresponding frequency variation.'' They found a null result,
i.e., a fractional length change of $\bigtriangleup l/l=(1.5\pm 2.5)\times
10^{-15}$ (this is also the fractional frequency shift) in showing the
isotropy of space; this result represented a 4000-fold improvement on the
measurements \cite{jaseja}. In \cite{haugan} the experiment \cite{brillet}
is quoted as the most precise repetition of the Michelson-Morley experiment,
and it is asserted that the experiment \cite{brillet} constrained the two
times, our $t_{1}^{\prime }$ and $t_{2}^{\prime }$, to be equal within a
fractional error of $10^{-15}$. The times $t_{1}^{\prime }$ and $%
t_{2}^{\prime }$ refer to the round-trips in two maser cavities in \cite
{jaseja}, and to the round-trips in the Fabry-Perot cavity in \cite{brillet}%
. These times are calculated in the same way as in the Michelson-Morley
experiment.(see, for example, \cite{haugan}).

The above brief discussion of the experiments \cite{jaseja} and \cite
{brillet}, and the previous analysis of the usual, ''AT relativity,''
calculation of $t_{1}^{\prime }$ and $t_{2}^{\prime }$ in the
Michelson-Morley experiment, suggest that the same remarks as in the
Michelson-Morley experiment hold also for the experiments \cite{jaseja} and 
\cite{brillet}. For example, the reflections of light in maser cavities or
in Fabry-Perot cavity happen on the moving mirrors as in the
Michelson-Morley experiment, which means that the optical paths between the
reflecting ends have to be calculated taking into account the Doppler
effect, i.e., as in Driscoll's procedure \cite{drisc}. In fact, the
interference of the light waves, e.g., the light waves with close
frequencies from two maser cavities in \cite{jaseja}, is always determined
by their phase difference and not only with their frequencies. Hence,
although the measurement of the beat frequency variation is more precise
than the measurement of the shift in the number of fringes, it cannot check
the validity of the ''AT relativity'' in the same measure as it can the
latter one. Also it has to be noted that the theoretical predictions for the
beat frequency variation are strongly dependent on the chosen
synchronization. Altogether, contrary to the generally accepted opinion, the
experiments \cite{jaseja} and \cite{brillet} do not confirm the validity of
the ''AT relativity.''

Regarding the ''TT relativity,'' the modern laser versions \cite{jaseja} and 
\cite{brillet} of the Michelson-Morley experiment are incomplete experiments
(only the beat frequency variation is measured) and cannot be compared with
the theory; in the ''TT relativity'' the same 4D quantity has to be
considered in relatively moving IFRs and \emph{the frequency, taken alone,
is not a 4D quantity}.

\section{THE KENNEDY-THORNDIKE TYPE EXPERIMENTS}

In the Kennedy-Thorndike experiment \cite{kenne} a Michelson interferometer
with unequal armlengths was employed and they looked for possible diurnal
and annual variations in the difference of the optical paths due to the
motion of the interferometer with respect to the preferred frame. The
measured quantity was, as in the Michelson-Morley experiment, the shift in
the number of fringes, and in \cite{kenne} the authors also found that was
no observable fringe shift. We shall not discuss this experiment since the
whole consideration is completely the same as in the case of the
Michelson-Morley experiment, and, consequently, the same conclusion holds
also here, i.e., the experiment \cite{kenne} does not agree with the ''AT
relativity,'' but directly proves the ''TT relativity.'' A modern version of
the Kennedy-Thorndike experiment was carried out in \cite{hils}, and the
authors stated: ''We have performed the physically equivalent measurement
(with the Kennedy-Thorndike experiment, my remark) by searching for a
sidereal 24-h variation in the frequency of a stabilized laser compared with
the frequency of a laser locked to a stable cavity.'' The result was: ''No
variations were found at the level of $2\times 10^{-13}."$ Also they
declared: ''This represents a 300-fold improvement over the original
Kennedy-Thorndike experiment and \emph{allows the Lorentz transformations to
be deduced entirely from experiment at an accuracy level of 70 ppm.'' }(my
emphasis) The experiment \cite{hils} is of the same type as the experiment 
\cite{brillet}, and neither the experiment \cite{brillet} is physically
equivalent to the Michelson-Morley experiment, as shown above, nor, contrary
to the opinion of the authors of \cite{hils}, the experiment \cite{hils} is
physically equivalent to the Kennedy-Thorndike experiment; the measurement
of the beat frequency variation is not equivalent to the measurement of the
change in the phase difference (in terms of the measurement of the shift in
the number of fringes). And, additionally, the Michelson-Morley and the
Kennedy-Thorndike experiments can be compared both with the ''AT
relativity'' and the ''TT relativity'', while the modern laser versions \cite
{brillet}, \cite{jaseja} and \cite{hils} of these experiments are incomplete
experiments from the ''TT relativity'' viewpoint and cannot be compared with
the ''TT relativity.'' Furthermore, the ''TT relativity'' deals with the
covariant 4D Lorentz transformations (Eq.(3) in $\left[ I\right] $ and in
Sec.3.3 here) and they cannot be deduced from the experiment \cite{hils}.

\section{THE IVES-STILLWEL TYPE EXPERIMENTS}

Ives and Stilwell \cite{ives} performed a precision Doppler effect
experiment in which they used a beam of excited hydrogen molecules as a
moving light source. The frequencies of the light emitted parallel and
antiparallel to the beam direction were measured by a spectograph (at rest
in the laboratory). The measured quantity in this experiment is 
\begin{equation}
\bigtriangleup f/f_{0}=(\bigtriangleup f_{b}-\bigtriangleup f_{r})/f_{0},
\label{frdop}
\end{equation}
where $f_{0}$ is the frequency of the light emitted from resting atoms. $%
\bigtriangleup f_{b}=\left| f_{b}-f_{0}\right| $ and $\bigtriangleup
f_{r}=\left| f_{r}-f_{0}\right| ,$ where $f_{b}$ is the blue-Doppler-shifted
frequency that is emitted in a direction parallel to $\mathbf{v}$ ($\mathbf{v%
}$ is the velocity of the atoms relative to the laboratory), and $f_{r}$ is
the red-Doppler-shifted frequency that is emitted in a direction opposite to 
$\mathbf{v.}$ The quantity $\bigtriangleup f/$ $f_{0}$ measures the extent
to which the frequency of the light from resting atoms fails to lie halfway
between the frequencies $f_{r}$ and $f_{b}.$ In terms of wavelengths the
relation (\ref{frdop}) can be written as 
\begin{equation}
\bigtriangleup \lambda /\lambda _{0}=(\bigtriangleup \lambda
_{r}-\bigtriangleup \lambda _{b})/\lambda _{0},  \label{ladop}
\end{equation}
where $\bigtriangleup \lambda _{r}=\left| \lambda _{r}-\lambda _{0}\right| $
and $\bigtriangleup \lambda _{b}=\left| \lambda _{b}-\lambda _{0}\right| ,$
and, as we said, $\lambda _{r}$ and $\lambda _{b}$ are the wavelengths
shifted due to the Doppler effect to the ''red'' and ''blue'' regions of the
spectrum. In that way Ives and Stilwell replaced the difficult problem of
the precise determination of the wavelength with much simpler problem of the
determination of the asymmetry of shifts of the ''red'' and ''blue'' shifted
lines with respect to the unshifted line. They \cite{ives} showed that the
measured results agree with the formula predicted by the traditional
formulation of the special relativity, i.e., the ''AT relativity,'' and not
with the classical nonrelativistic expression for the Doppler effect. Let us
explain it in more detail.

\subsection{The ''AT relativity'' calculation}

In the ''AT relativity'' one usually starts with the Lorentz transformation
of the 4-vector $k^{\mu }(\omega /c,\mathbf{k=n}\omega /c\mathbf{)}$ of the
light wave from an IFR $S$ to the relatively moving (along the common $%
x,x^{\prime }-$axes) IFR $S^{\prime }$. Note that only the ''e''
coordinatization is used in such traditional treatment. Then the Lorentz
transformation in the ''e'' base of $k^{\mu }$ can be written as 
\begin{equation}
\omega ^{\prime }/c=\gamma (\omega /c-\beta k^{1}),k^{\prime 1}=\gamma
(k^{1}-\beta \omega /c),k^{\prime 2}=k^{2},k^{\prime 3}=k^{3},
\label{lordop}
\end{equation}
or in terms of the unit wave vector $\mathbf{n}$ (which is in the direction
of propagation of the wave) 
\begin{equation}
\omega ^{\prime }=\gamma \omega (1-\beta n^{1}),n^{\prime 1}=N(n^{1}-\beta
),n^{\prime 2}=(N/\gamma )n^{2},n^{\prime 3}=(N/\gamma )n^{3},  \label{odop}
\end{equation}
where $N=(1-\beta n^{1})^{-1}.$ Now comes the main point in the derivation.
Although the Lorentz transformation of the 4-vector $k^{\mu }$ from $S$ to $%
S^{\prime },$ Eqs.(\ref{lordop}) and (\ref{odop}), transforms all four
components of $k^{\mu }$ the usual ''AT relativity'' treatment considers the
transformation of the temporal part of $k^{\mu },$ i.e., the frequency, as
independent of the transformation of the spatial part of $k^{\mu },$ i.e.,
the unit wave vector $\mathbf{n,}$ and thus the ''AT relativity'' deals with
two independent physical phenomena - the Doppler effect and the aberration
of light. We note once again that such distinction is possible only in the
''e'' coordinatization; in the ''r'' base the metric tensor is not diagonal
and consequently the separation of the temporal and spatial parts does not
exist. Thus the ''AT relativity'' calculation is restricted to the ''e''
base. In agreement with such theoretical treatment the existing experiments
(including the modern experiments based on collinear laser spectroscopy;
see, e.g., \cite{mcgow,riis,klein}, or the review \cite{kretz}) are designed
in such a way to measure either the Doppler effect or the aberration of
light. Let us write the above transformation in the form from which one can
determine the quantities in (\ref{ladop}) and then compare with the
experiments. The spectograph is at rest in the laboratory (the $S$ frame)
and the light source (at rest in the $S^{\prime }$ frame) is moving with $%
\mathbf{v}$ relative to $S.$ Then in the usual ''AT relativity'' approach 
\emph{only the first relation from (\ref{lordop}), or (\ref{odop}), is used,}
which means that, in the same way as shown in previous cases, \emph{the ''AT
relativity'' deals with two different quantities in 4D spacetime, here }$%
\omega $\emph{\ and }$\omega ^{\prime }$\emph{.} Then writting the
transformation of the temporal part of $k^{\mu },$ i.e., of $\omega ,$ in
terms of the wavelength $\lambda $ we find 
\begin{equation}
\lambda =\gamma \lambda _{0}(1-\beta \cos \theta ),  \label{lam1}
\end{equation}
where $\lambda $ is the wavelength received in the laboratory from the
moving source (the shifted line), $\lambda _{0}$ ($=\lambda ^{\prime }$) is
the natural wavelength (the unshifted line) and $\theta $ is the angle of $%
\mathbf{k}$ relative to the direction of $\mathbf{v}$ as measured in the
laboratory. The nonrelativistic treatment of the Doppler effect predicts $%
\lambda =\lambda _{0}(1-\beta \cos \theta ),$ and in the classical case the
Doppler shift does not exist for $\theta =\pi /2$. This transverse Doppler
effect ($\theta =\pi /2,$ $\lambda =\gamma \lambda _{0},$ or $\nu =\nu
_{0}/\gamma $) is always, in the traditional, ''AT relativity,'' approach
considered to be a direct consequence of time dilatation; it is asserted
(e.g. \cite{Kittel}) that the frequencies must be related as the inverse of
the times in the usual relation for the time dilatation $\bigtriangleup
t=\bigtriangleup t_{0}\gamma $. It is usually interpreted \cite{kretz}:
''The Doppler shift experiments ... compare the rates of two ''clocks'' that
are in motion relative to each other. \emph{They measure time dilatation}
(my emphasis) and can test the validity of the special relativity in this
respect.'' Similarly it is declared in \cite{mcgow}: ''The experiment
represents a more than tenfold improvement over other Doppler shift
measurements and \emph{verifies the time dilation effect} (my emphasis) at
an accuracy level of 2.3ppm.'' Obviously, as we said, the Doppler shift
experiments are theoretically analysed only by means of the ''AT
relativity,'' which treats the transformation of the temporal part of $%
k^{\mu }$ as independent of the transformation of the spatial part of $%
k^{\mu }$.

In the Ives and Stilwell type experiments the measurements are conducted at
symmetric observation angles $\theta $ and $\theta +180^{0};$ particularly
in \cite{ives} $\theta $ is chosen to be $\simeq 0^{0}$. The wavelength in
the direction of motion is obtained from (\ref{lam1}) as $\lambda
_{b}=\gamma \lambda _{0}(1-\beta \cos \theta ),$ while that one in the
opposite direction (the angle $\theta +180^{0}$) is $\lambda _{r}=\gamma
\lambda _{0}(1+\beta \cos \theta ),$ and then $\bigtriangleup \lambda
_{b}=\left| \lambda _{b}-\lambda _{0}\right| =\left| \lambda _{0}(1-\gamma
+\beta \gamma \cos \theta )\right| ,$ $\bigtriangleup \lambda _{r}=\left|
\lambda _{r}-\lambda _{0}\right| =\left| \lambda _{0}(\gamma -1+\beta \gamma
\cos \theta )\right| ,$ and the difference in shifts is 
\begin{equation}
\bigtriangleup \lambda =\bigtriangleup \lambda _{r}-\bigtriangleup \lambda
_{b}=2\lambda _{0}(\gamma -1)\simeq \lambda _{0}\beta ^{2},  \label{dellam}
\end{equation}
where the last relation holds for $\beta \ll 1.$ Note that the redshift due
to the transverse Doppler effect ($\lambda _{0}\beta ^{2}$) is independent
on the observation angle $\theta $. In the nonrelativistic case $%
\bigtriangleup \lambda =0$, the transverse Doppler shift is zero. Ives and
Stilwell found the agreement of the experimental results with the relation (%
\ref{dellam}) and not with the classical result $\bigtriangleup \lambda =0.$

However, a more careful analysis shows that the agreement between the ''AT
relativity'' prediction Eq.(\ref{dellam}) and the experiments \cite{ives}
is, contrary to the general belief, only an ''apparent'' agreement and not
the ''true'' one. This agreement actually happens for the following reasons.
First, the theoretical result (\ref{dellam}) is obtained in the ''e''
coordinatization in which one can speak about the frequency $\omega $ and
the wave vector $\mathbf{k}$ as well-defined quantities. Using the matrix $%
T_{\nu }^{\mu },$ which transforms the ''e'' coordinatization to the ''r''
coordinatization, $k_{r}^{\mu }=T_{\nu }^{\mu }k_{e}^{\nu },$ (see, $\left[
I\right] $) one finds $k_{r}^{0}=k_{e}^{0}-k_{e}^{1}-k_{e}^{2}-k_{e}^{3},%
\quad k_{r}^{i}=k_{e}^{i},$ whence we conclude that in the ''r'' base the
theoretical predictions for \emph{the components} of a 4-vector, i.e., for $%
\lambda ,$ will be quite different than in the ''e'' base, i.e., than the
result (\ref{dellam}), and thus not in the agreemement with the experiment 
\cite{ives}. Further, the specific choice of $\theta $ ($\theta \simeq
0^{0}) $ in the experiments \cite{ives} is the next reason for the agreement
with the ''AT relativity'' result (\ref{dellam}). Namely, if $\theta =0^{0}$
than $n^{1}=1,$ $n^{2}=n^{3}=0$, and $k^{\mu }$ is $(\omega /c,\omega /c,0,0%
\mathbf{).}$ From (\ref{lordop}) or (\ref{odop}) one finds that in $%
S^{\prime }$ too $\theta ^{\prime }=0^{0},$ $n^{\prime 1}=1$ and $n^{\prime
2}=n^{\prime 3}=0$ (the same holds for $\theta =180^{0},$ $n^{1}=-1,$ then $%
\theta ^{\prime }=180^{0}$ and $n^{\prime 1}=-1$). In the experiments \cite
{ives} the emitter is the moving ion (its rest frame is $S^{\prime }$),
while the observer is the spectrometer at rest in the laboratory (the $S$
frame). Since in \cite{ives} the angle of the ray emitted by the ion at rest
is chosen to be $\theta ^{\prime }=0^{0}$ ($180^{0}$), then the angle of
this ray measured in the laboratory, where the ion is moving, will be the
same $\theta =0^{0}$ ($180^{0}$). (Similarly happens in the modern versions 
\cite{mcgow,klein} of the Ives-Stilwell experiment; the experiments \cite
{mcgow,klein} make use of an atomic or ionic beam as a moving light analyzer
(the accelerated ion is the ''observer'') and two collinear laser beams
(parallel and antiparallel to the particle beam) as light sources (the
emitter), which are at rest in the laboratory.) From this consideration we
conclude that in these experiments one can consider only the Doppler effect,
that is, the transformation of $\omega $ (the temporal part of $k^{\mu };$ $%
k^{a}$ in the ''e'' base), and not the aberration of light, i.e., the
transformation of $\mathbf{n,}$ i.e., $\mathbf{k,}$ (the spatial part of $%
k^{\mu }$); because of that they found the agreement between the relation (%
\ref{lam1}) (or (\ref{dellam})) with the experiments. However, the relations
(\ref{lordop}) and (\ref{odop}) reveal that in the case of an arbitrary $%
\theta $ the transformation of the temporal part of $k^{\mu }$ cannot be
considered as independent of the transformation of the spatial part, which
means that in this case one cannot expect that the relation (\ref{dellam}),
taken alone, will be in agreement with the experiments performed at some
arbitrary $\theta .$ Such experiments were, in fact, recently conducted and
we discuss them here.

Pobedonostsev and collaborators \cite{pobed} performed the Ives-Stilwell
type experiment but improved the experimental setup and, what is
particularly important, the measurements were conducted at symmetric
observation angles $77^{0}$ and $257^{0}.$ The work was done with a beam of $%
H_{2}^{+}$ ions at energies $175,180,210,225,260$ and $275$ $keV,$ and the
radiation from hydrogen atoms in excited state, which are formed as a result
of disintegration of accelerated $H_{2}^{+},$ was observed. The radiation
from the moving hydrogen atoms, giving the Doppler shifted lines, was
observed together with the radiation from the resting atoms existing in the
same working volume, and giving an unshifted line. The similar work was
reported in \cite{pobe2} in which a beam of $H_{3}^{+}$ ions at energy $310$ 
$keV$ was used and the measurements were conducted at symmetric observation
angles $82^{0}$ and $262^{0}.$ The results of the experiments \cite{pobed}
and \cite{pobe2} markedly differed from all previous experiments that were
performed at observation angles $\theta =0^{0}$ (and $180^{0}$). Therefore
in \cite{pobe2} Pobedonostsev declared: \emph{''In comparing the wavelength
of Doppler shifted line from a moving emitter with the wavelength of an
identical static emitter, the experimental data corroborate the classical
formula for the Doppler effect, not the relativistic one.''} Thus, instead
of to find the ''relativistic'' result $\bigtriangleup \lambda \simeq
\lambda _{0}\beta ^{2}$ (\ref{dellam}), (actually the ''AT relativity''
result), they found the classical result $\bigtriangleup \lambda \simeq 0,$
i.e., they found that the redshift due to the transverse Doppler effect ($%
\lambda _{0}\beta ^{2}$) \emph{is dependent} on the observation angle $%
\theta $. This experimental result strongly support our assertion that the
agreement between the ''AT relativity'' and the Ives-Stilwell type
experiments is only an ''apparent'' agreement and not the ''true'' one.

\subsection{The ''TT relativity'' approach}

As already said in the ''TT relativity'' neither the Doppler effect nor the
aberration of light exist separately as well defined physical phenomena. As
shown in Sec.3.3 here and in $\left[ I\right] ,$ in 4D spacetime the
temporal distances (e.g., $\tau _{E}$ and $\tau _{\mu }$) refer to different
quantities, which are not connected by the Lorentz transformation; the same
happens with $\omega $ and $\omega ^{\prime }$ as the temporal parts of $%
k^{\mu }$. And, as Gamba \cite{gamba} stated, the fact that the measurements
of such quantities were made by \emph{two} observers does not mean that
relativity has something to do with the problem. In the ''TT relativity''
the entire 4D quantity has to be considered both in the theory and in
experiments. Therefore, in order to theoretically discuss the experiments of
the Ives-Stilwell type we choose as the relevant quantity the wave vector $%
k^{a}$ and its square, for which it holds that 
\begin{equation}
k^{a}g_{ab}k^{b}=0.  \label{kadop}
\end{equation}
First we consider the experiments \cite{pobed} and \cite{pobe2} since they
showed the disagreement with the traditional theory, i.e., with the ''AT
relativity.'' The product $k^{a}g_{ab}k^{b}$ is a Lorentz scalar and it is
also independent of the choice of the coordinatization. Hence we can
calculate that product $k^{a}g_{ab}k^{b}$ in the ''e'' base and in the rest
frame of the emitter (the $S^{\prime }$ frame); the emitter is the ion
moving in $S,$ the rest frame of the spectrometer, i.e., in the laboratory
frame. Then $k^{a}$ in the ''e'' base and in $S^{\prime }$ is represented by 
$k^{\prime \mu }=(\omega ^{\prime }/c)(1,\cos \theta ^{\prime },\sin \theta
^{\prime },0\mathbf{)}$ and $k^{\prime \mu }k_{\mu }^{\prime }=0.$ The
observer (the spectrometer) in the laboratory frame will look at \emph{the
same 4D quantity} $k^{a}$ and find the Lorentz transformed wave vector $%
k^{\mu }$ as 
\[
k^{\mu }=\left[ \gamma (\omega ^{\prime }/c)(1+\beta \cos \theta ^{\prime
}),\gamma (\omega ^{\prime }/c)(\cos \theta ^{\prime }+\beta ),(\omega
^{\prime }/c)\sin \theta ^{\prime },0\right] , 
\]
whence $k^{\mu }k_{\mu }$ is also $=0.$ From that transformation one can
find that 
\[
n^{1}=(n^{\prime 1}+\beta )/(1+\beta n^{\prime 1}),n^{2}=n^{\prime 2}/\gamma
(1+\beta n^{\prime 1}),n^{3}=n^{\prime 3}/\gamma (1+\beta n^{\prime 1}), 
\]
or that 
\[
\sin \theta =\sin \theta ^{\prime }/\gamma (1+\beta \cos \theta ^{\prime
}),\cos \theta =(\cos \theta ^{\prime }+\beta )/(1+\beta \cos \theta
^{\prime }), 
\]
\begin{equation}
\tan \theta =\sin \theta ^{\prime }/\gamma (\beta +\cos \theta ^{\prime }).
\label{tanab}
\end{equation}
The relation (\ref{tanab}) reveals that not only $\omega $ is changed (the
Doppler effect) when going from $S^{\prime }$ to $S$ than also the angle of $%
\mathbf{k}$ relative to the direction of $\mathbf{v}$ is changed (the
aberration of light). This means that if the observation of the unshifted
line (i.e., of the frequency $\omega ^{\prime }=\omega _{0}$ from the atom
at rest) is performed at an observation angle $\theta ^{\prime }$ in $%
S^{\prime },$ the rest frame of the emitter, then \emph{the same light wave}
(from the same but now moving atom) will have the shifted frequency $\omega $
and will be seen at an observation angle $\theta $ (generally, $\neq \theta
^{\prime }$) in $S,$ the rest frame of the spectrometer. In $S^{\prime }$
the quantities $\omega ^{\prime }$ and $\theta ^{\prime }$ define $k^{\prime
\mu },$ and this propagation 4-vector satisfies the relation $k^{\prime \mu
}k_{\mu }^{\prime }=0,$ which is the representation of the relation (\ref
{kadop}) in the ''e'' base and in the $S^{\prime }$ frame. The quantities $%
\omega ^{\prime }$ and $\theta ^{\prime }$ are connected with the
corresponding $\omega $ and $\theta ,$ that define the corresponding $k^{\mu
}$ in $S,$ by the ''e'' base Lorentz transformation of $k^{\prime \mu }.$
Then $k^{\mu }$ is such that it also satisfies the relation $k^{\mu }k_{\mu
}=0$ (the representation of the same relation (\ref{kadop}) in the ''e''
base and now in the $S$ frame). The authors of the experiments \cite{pobed}
(and \cite{pobe2}) made the observation of the radiation from the atom at
rest (the unshifted line) and from a moving atom at the same observation
angle. The preceding discussion shows that if they succeeded to see $\omega
_{0}$ ($\lambda _{0}$) from the atom at rest at some symmetric observation
angles $\theta ^{\prime }$ ($\neq 0$) and $\theta ^{\prime }+180^{0}$ than
they could not see the assymetric Doppler shift (from moving atoms) at the
same angles $\theta =\theta ^{\prime }$ (and $\theta ^{\prime }+180^{0}$).
This was the reason that they detected $\bigtriangleup \lambda \simeq 0$ and
not $\bigtriangleup \lambda \simeq \lambda _{0}\beta ^{2}.$ But we expect
that the result $\bigtriangleup \lambda \simeq \lambda _{0}\beta ^{2}$ can
be seen if the similar measurements of the frequencies, i.e., the
wavelengths, of the radiation from moving atoms would be performed not at $%
\theta =\theta ^{\prime }$ than at $\theta $ determined by the relation (\ref
{tanab}).

Recently, Bekljamishev \cite{bekla} came to the same conclusions and
explained the results of the experiments \cite{pobed} and \cite{pobe2}
taking into account the aberration of light together with the Doppler
effect. It is argued in \cite{bekla} that Eq.(\ref{lam1}) for the Doppler
effect can be realized only when the condition for the aberration angle is
fulfilled, 
\begin{equation}
\bigtriangleup \theta =\beta \sin \theta ^{\prime },  \label{aber}
\end{equation}
where $\bigtriangleup \theta =\theta ^{\prime }-\theta ,$ and $\beta $ is
taken to be $\beta \ll 1.$ The relation (\ref{aber}) directly follows from
the expression for $\sin \theta $ in (\ref{tanab}) taking that $\beta \ll 1.$
The assymetric shift will be seen when the collimator assembly is tilted at
a velocity dependent angle $\bigtriangleup \theta .$ Instead of to work, as
usual, with the arms of the collimator at fixed angles $\theta $ and $\theta
+180^{0},$ Bekljamishev \cite{bekla} proposed that the collimator assembly
must be constructed in such a way that there is the possibility of the
correction of the observation angles independently for both arms; for
example, the arm at angle $\theta $ ($\theta +180^{0}$) has to be tilted
clockwise (counter-clockwise) by the aberration angle $\bigtriangleup \theta
.$ Otherwise the assymetry in the Doppler shifts will not be observed. Thus
the experiments \cite{pobed} and \cite{pobe2} would need to be repeated
taking into account Bekljamishev's proposition. The positive result for the
Doppler shift $\bigtriangleup \lambda $ (\ref{dellam}), when the condition
for the aberration angle $\bigtriangleup \theta $ (\ref{aber}) is fulfilled,
will definitely show that it is not possible to treat the Doppler effect and
the aberration of light as separate, well-defined, effects, i.e., that it is
the ''TT relativity,'' and not the ''AT relativity,'' which correctly
explains the experiments that test the special relativity.

\section{CONCLUSIONS\ AND\ DISCUSSION}

The analysis of the experiments which test the special relativity shows that
they agree with the predictions of the ''TT relativity'' and not, as usually
supposed, with those of the ''AT relativity.''

In the ''muon'' experiment the fluxes of muons on a mountain, $N_{m}$, and
at sea level, $N_{s}$, are measured. The ''AT relativity'' predicts
different values of the flux $N_{s}$ (for the same measured $N_{m}$) in
different synchronizations, but the measured $N_{s}$ is independent of the
chosen coordinatization. Further, for some synchronizations these predicted
values of the flux at sea level $N_{s}$ are quite different than the
measured ones. The reason for such disagreement, as explained in the text,
is that in the usual, ''AT relativity,'' analysis of the ''muon''
experiment, for example, the lifetimes $\tau _{E}$ and $\tau _{\mu }$ are
considered to refer to the same temporal distance (the same quantity)
measured by the observers in two relatively moving IFRs. But the
transformation connecting $\tau _{E}$ and $\tau _{\mu }$ (the dilatation of
time) is only \emph{a part} of the Lorentz transformation written in the
''e'' base, and, actually, $\tau _{E}$ and $\tau _{\mu }$ refer to different
quantities in 4D spacetime. Although their measurements were made by\emph{\
two} observers, the relativity has nothing to do with the problem, since $%
\tau _{E}$ and $\tau _{\mu }$ are different 4D quantities. \emph{The ''TT
relativity,'' in contrast to the ''AT relativity,'' completely agrees with
the ''muon'' experiments in all IFRs and all possible coordinatizations. }In
the ''TT relativity''\emph{\ the same 4D quantity} is considered in
different IFRs and different coordinatizations; instead of to work with $%
\tau _{E}$ and $\tau _{\mu }$ the ''TT relativity'' deals with the spacetime
length $l$ and formulate the radioactive-decay law in terms of covariantly
defined quantities.

In the Michelson-Morley experiment the traditional, ''AT relativity,''
derivation of the fring shift $\bigtriangleup N$ deals only with the
calculation, in the ''e'' base, of path lengths (optical or geometrical) in $%
S$ and $S^{\prime },$ or, in other words, with the calculation of $t_{1}$
and $t_{2}$ (in $S$ and $S^{\prime }$)$,$ which are the times required for
the complete trips $OM_{1}O$ and $OM_{2}O$ along the arms of the
Michelson-Morley interferometer. The null fringe shift obtained with such
calculation is only in an ''apparent,'' not ''true,'' agreement with the
observed null fringe shift, since this agreement was obtained by an
incorrect procedure. Namely it is supposed in such derivation that, e.g., $%
t_{1}$ and $t_{1}^{\prime }$ refer to the same quantity measured by the
observers in relatively moving IFRs $S$ and $S^{\prime }$ that are connected
by the Lorentz transformation. However the relation $t_{1}^{\prime }=\gamma
t_{1}$ is not the Lorentz transformation of some 4D quantity, and $%
t_{1}^{\prime }$\ and $t_{1}$\ do not correspond to the same 4D quantity
considered in $S^{\prime }$ and $S$ respectively. The improved ''AT
relativity'' calculation from \cite{drisc} (again in the ''e'' base)
determines the increment of phase along some path not only by the segment of
geometric path length than also by the wavelength in that segment, and finds
a non-null fringe shift. But we show that the non-null theoretical result
for the fringe shift, which is obtained in \cite{drisc}, is a consequence of
the fact that again two different quantities $k_{e}^{0}l_{0e}$ and $%
k_{e}^{\prime 0}l_{0e}^{\prime }$ (only the parts of the covariantly defined
phase (\ref{phase}) $\phi =k^{a}g_{ab}l^{b}$) are considered to refer to the
same 4D quantity, and thus that these two quantities are connected by the
Lorentz transformation. The ''TT relativity,'' in contrast to the ''AT
relativity'' calculations, deals always with the covariantly defined 4D
quantities (in the Michelson-Morley experiment, e.g., the covariantly
defined phase (\ref{phase}) $\phi =k^{a}g_{ab}l^{b}$), which \emph{are}
connected by the Lorentz transformation. \emph{The ''TT relativity''
calculations yields the observed null fringe shift and that result holds for
all IFRs and all coordinatizations.}

The same conclusions can be drawn for the Kennedy-Thorndike type experiments.

In the Ives-Stilwell type experiments the agreement between the ''AT
relativity'' calculation for the Doppler effect and the experiments is again
only an ''apparent'' agreement and not the ''true'' one. Namely the
transverse Doppler shift ($\lambda _{0}\beta ^{2}$, (\ref{dellam})) is
obtained in the ''e'' coordinatization in which one can speak about the
frequency $\omega $ and the wave vector $\mathbf{k}$ as well-defined
quantities. Further in the usual ''AT relativity'' approach only the
transformation of $\omega $ (the temporal part of $k^{\mu }$) is considered,
while the aberration of light, i.e., the transformation of $\mathbf{n,}$
i.e., $\mathbf{k,}$ (the spatial part of $k^{\mu }$) is neglected. Thus in
this case too the ''AT relativity'' deals with two different quantities in
4D spacetime, $\omega $\ and $\omega ^{\prime }$, which are not connected by
the Lorentz transformation. However, for the specific choice of the
observation angles $\theta ^{\prime }=0^{0}$ ($180^{0}$) in $S^{\prime }$
(the rest frame of the emitter), one finds from the transformation of $%
k^{\mu }$ that $\theta $ in $S$ is again $=0^{0}$ ($180^{0}$). Since in the
experiments \cite{ives}, and its modern versions \cite{mcgow,klein}, just
such angles were chosen, it was possible to consider only the transformation
of $\omega $, i.e., only the Doppler effect, and not the concomitant
aberration of light, and because of that they found the agreement between
the relation (\ref{lam1}) (or (\ref{dellam})) and the experiments. When the
experiments were performed at observation angles $\theta \neq 0^{0}$ (and $%
180^{0}$), as in \cite{pobed} and \cite{pobe2}, the results disagreed with
the ''AT relativity'' calculation which takes into account only the
transformation of $\omega $, i.e., only the Doppler effect. The ''TT
relativity'' calculation considers \emph{the same 4D quantity} the wave
vector $k^{a}$ (or its square) in $S$ and $S^{\prime },$ i.e., it considers
the Doppler effect and the aberration of light together as unseparated
phenomena. The results of such calculation agrees with the experiments \cite
{ives} and \cite{mcgow,klein} (made at $\theta =0^{0}$ ($180^{0}$)), but
also predict the positive result for the Doppler shift $\bigtriangleup
\lambda $ (\ref{dellam}) in the experiments of the type \cite{pobed} and 
\cite{pobe2}, if the condition for the aberration angle $\bigtriangleup
\theta $ (\ref{aber}) is fulfilled, which agrees with Bekljamishev's
explanation \cite{bekla} of the experiments \cite{pobed} and \cite{pobe2}.

The discussion in this paper clearly shows that \emph{the ''TT relativity''
completely agrees with all considered experiments, in all IFRs and all
possible coordinatizations,} while the ''AT relativity'' appears as an
unsatisfactory relativistic theory. These results are directly contrary to
the generally accepted opinion about the validity of the ''AT relativity.''

\end{document}